\documentclass[12pt]{article}
\usepackage{amsfonts}
\usepackage{graphicx}
\usepackage[nosort]{cite}
\usepackage{amssymb,amsmath,amsthm}
\usepackage{enumerate}
\usepackage{mathrsfs}
\usepackage[dvipsnames]{xcolor}
\usepackage{mathtools}
\usepackage{booktabs}
\usepackage{array}
\usepackage{verbatim}
\usepackage{empheq}

\usepackage[english]{babel}
\usepackage{braket}
\usepackage{bm}
\usepackage{float}
\usepackage{appendix}

\usepackage{longtable}
\usepackage{tikz}
\usetikzlibrary{decorations.markings,arrows.meta,shapes.geometric}
\tikzset{fermion/.style={thick,postaction={decorate},decoration={markings,mark=at position 0.56 with {\arrow{Stealth[length=5pt,width=4pt]}}}},
  scalar/.style={thick,dashed,dash pattern=on 3pt off 2pt}, vlab/.style={font=\scriptsize,text=black!55,inner sep=1.5pt}}
\newcommand{\Tpsi}[2]{\filldraw[thick,fill=white] (#1,#2) circle (0.17);
  \draw[thick] (#1,#2) ++(-0.12,-0.12) -- ++(0.24,0.24) (#1,#2) ++(-0.12,0.12) -- ++(0.24,-0.24);}
\newcommand{\Tsig}[2]{\filldraw[thick,fill=white] (#1,#2) ++(-0.16,-0.16) rectangle ++(0.32,0.32);
  \draw[thick] (#1,#2) ++(-0.16,-0.16) -- ++(0.32,0.32) (#1,#2) ++(-0.16,0.16) -- ++(0.32,-0.32);}
\newcommand{\eps}{\epsilon}

\usepackage[hidelinks]{hyperref}

\makeatletter\@addtoreset{equation}{section}\makeatother

\renewcommand{\title}[1]{\vbox{\center\LARGE{#1}}\vspace{5mm}}
\renewcommand{\author}[1]{\vbox{\center#1}\vspace{5mm}}
\newcommand{\address}[1]{\vbox{\center\em#1}}

\newcommand{\Tr}{\mathrm{Tr}}  % fermion-loop trace, as used throughout this paper

\begin{document}
\newtheorem{theorem}{Theorem}
\newtheorem{lemma}{Lemma}
\newtheorem{conjecture}{Conjecture}
\newtheorem{example}{Example}
\newtheorem{corollary}{Corollary}

\hypersetup{pageanchor=false}
\begin{titlepage}
\begin{center}

\vskip 1cm

\setcounter{tocdepth}{1}

\title{$C_T$ and $C_J$ of general scalar--fermion conformal field theories in $d=4-\epsilon$}

\author{Shiroman Prakash and Shubham Kumar Sinha}

\address{Department of Physics and Computer Science, Dayalbagh Educational Institute, Agra, India}

\end{center}

\begin{abstract}
We compute $C_T$ and $C_J$, the coefficients of the stress-tensor and conserved-current two-point functions, to order $\epsilon^3$ at any fixed
point of the general renormalizable theory of scalars and fermions in $d=4-\epsilon$.  We give explicit results for the chiral Ising, XY and
Heisenberg models and for fixed points with emergent supersymmetry in $d=3$.
In theories without a four-dimensional counterpart, fermion loops with an odd number of Yukawa vertices contribute at order $\epsilon^3$ through
traces whose continuation away from $d=3$ is ambiguous; we fix it with $\mathcal N=1$ supersymmetry, by computing the two-point functions of the
supercurrent and of the fermionic partner of a flavor current to four loops.  In $d=3$, Pad\'e approximants closely reproduce the exact $C_T$ of
the $\mathcal N=2$ super-Ising model.  We estimate the universal conductivity at the charge-density-wave, Kekul\'e and N\'eel transitions of
graphene to be $\sigma/\sigma_{\rm free}\approx0.91$, $0.79$--$0.83$ and $0.64$--$0.74$, and show that the exact conductivity at the
supersymmetric transition on the surface of a topological insulator is $\sigma/\sigma_{\rm free}=\frac{9}{10}\,C_T/C_{T,\rm free}\approx0.654$,
$10\%$ below the value previously reported.
\end{abstract}

\vfill

\end{titlepage}

\eject \hypersetup{pageanchor=true}\tableofcontents

\clearpage

\section{Introduction}
We study a general renormalizable quantum field theory in four dimensions containing scalars and spin-$\frac12$ fermions,
interacting through Yukawa couplings $y^a$ and quartic couplings $\lambda_{abcd}$. In $d=4-\epsilon$ both couplings become weakly relevant, and the
renormalization-group flow can end at fixed points where $y_*^2$ and $\lambda_*$ are of order $\epsilon$, just as in the Wilson--Fisher fixed point of
scalar theories with quartic interactions \cite{WilsonFisher}.  The conformal data of these fixed points can then be computed order by order in $\epsilon$ and extrapolated to $d=3$.
The simplest examples are the Gross--Neveu model \cite{GrossNeveu} in its Gross--Neveu--Yukawa (GNY) form \cite{ZinnJustin}, also called the chiral Ising
model, and its chiral XY and Heisenberg cousins.  In $2+1$ dimensions they describe quantum phase transitions of Dirac fermions, such as the
semimetal--insulator transitions of electrons on the honeycomb lattice of graphene \cite{Herbut2006,HerbutJuricicVafek}. 

The chiral models have been studied in the $\epsilon$-expansion up to five loops \cite{Mihaila,Zerf,IhrigMihailaScherer,GraceyFiveLoop} and with the
conformal bootstrap \cite{Iliesiu,IliesiuGlobal,Erramilli,MitchellPoland,RongSu,Atanasov}. For the general theory, with arbitrary $y^a$ and
$\lambda_{abcd}$, the $\beta$-functions have been studied at two loops \cite{MV1,MV2,MV3,Luo}, at three loops
\cite{Steudtner,BednyakovPikelner,DaviesHerrenThomsen,HerrenMihailaSteinhauser, JOS}, and at four loops \cite{Steudtner2025}.  The fixed points of the general theory have been
surveyed in \cite{JOS,PannellStergiou,LiendoRong}, and they include many beyond the chiral models, some of them supersymmetric \cite{LiendoRong}.
Such fixed points may arise at critical points
of lattice models and on the boundaries of topological phases \cite{Lee2007,PonteLee,GroverShengVishwanath,FGKT,LiendoRong}.

In any conformal field theory, the two-point functions of the stress tensor and of a conserved current are fixed up to two coefficients, $C_T$ and $C_J$,
\begin{equation*}
\langle T_{\mu\nu}(x)T_{\rho\sigma}(0)\rangle=C_T\,\frac{I_{\mu\nu,\rho\sigma}(x)}{x^{2d}},\qquad
\langle J_\mu(x)J_\nu(0)\rangle = - C_J\,\frac{I_{\mu\nu}(x)}{x^{2(d-1)}},
\end{equation*}
where $I_{\mu\nu}$ and $I_{\mu\nu,\rho\sigma}$ are the standard inversion tensors \cite{OsbornPetkou}, given in Appendix \ref{app:method}. The minus sign arises because we take $J_\mu$ to be the Noether current divided by $i$ (section~\ref{sec:method}), so that $C_J>0$.
They enter the conformal bootstrap directly, since they fix the contributions of $T$ and $J$ to four-point functions.
$C_T$ generalizes the central charge of two-dimensional CFTs, and in $d=4$ it is proportional to the Weyl-anomaly coefficient, $c$.
$C_J$ also has physical significance -- in $2+1$ dimensions, $C_J$ of a $U(1)$ current determines the universal zero-temperature conductivity at the
critical point \cite{HuhStrackSachdev,HuhStrack,WitczakKrempa}.  At $\mathcal N=2$ fixed points, $C_T$ and $C_J$ of the $R$-current are related by
supersymmetry and are known exactly from localization \cite{NishiokaYonekura,WitczakKrempa,Bobev}, which makes them a non-trivial test of both
the $\epsilon$-expansion and emergent supersymmetry.

For the $O(N)$ model, consisting only of scalars, $C_T$ and $C_J$ are known at large $N$ \cite{Petkou} and in the $\epsilon$-expansion to $\mathcal O(\epsilon^4)$ \cite{GKSS,DKS,HvL},
and for a single scalar $C_T$ has also been computed directly at four loops \cite{IK}. For general renormalizable theories in four dimensions, corrections to $c$ are known to leading order  \cite{Hathrell1,Hathrell2,JackOsborn}, and they have been used to obtain $C_T$ at fixed points in $d=4-\epsilon$ \cite{OS2016}.  For the Gross--Neveu model, the leading large-$N$ corrections and the $\mathcal O(\epsilon)$ terms in $d=4-\epsilon$ were obtained in
\cite{Diab}, and extended to the chiral XY model in \cite{FGKT}.  However, to the best of our knowledge, beyond these leading terms, no higher-loop results for $C_T$ and $C_J$ were available for general scalar--fermion theories.

In this paper, we compute $C_T$ and $C_J$  up to $\mathcal O(\epsilon^3)$ at any fixed point of the general theory. Reaching $\mathcal O(\epsilon^3)$ requires four-loop diagrams. But we need only their pole terms, up to $1/\epsilon$, because the Fourier transform to position space gives an extra factor of $\epsilon$ (section~\ref{sec:renorm}).  At a fixed point the results depend only on coupling-tensor invariants
with $\epsilon$-independent coefficients, so they apply to any fixed point once its couplings are known.   The results pass a range of independent checks against large-$N$, $O(n)$,
analytic-bootstrap and supersymmetry results, and against a partially independent computation of the finite-$N$ Gross--Neveu--Yukawa result
(appendix~\ref{sec:checks}).

Our main results are the general formulas for $C_T$ and $C_J$, given in equations \eqref{eq:CT} and \eqref{eq:CJ} up to $\mathcal O(\epsilon^2)$, and in equations \eqref{eq:CT3} and \eqref{eq:CJ3} up to
$\mathcal O(\epsilon^3)$.  They hold for every such theory, with or without a four-dimensional counterpart. Their last terms, which we explain
next, contribute only to three-dimensional theories without one, such as a single two-component Majorana fermion coupled to a real scalar.  Every leading correction enters with a negative sign, so these results imply that, for real couplings, as in the $4-\epsilon$ continuation of a unitary theory, $C_T$ and $C_J$ lie below their free
values at the first order at which they differ from them.

These last terms come from fermion loops with an odd number of Yukawa vertices.  They vanish for every theory obtained from four dimensions, but
not for three-dimensional theories without a four-dimensional counterpart, such as the $\mathcal N=1$ fixed points and the matrix models.  There
they involve traces of five $\gamma$-matrices, whose continuation away from $d=3$ is not fixed by dimensional regularization (section~\ref{sec:dirac}).
We fix it with $\mathcal N=1$ supersymmetry, by computing the two-point functions of the supercurrent and of the fermionic partner $\theta_{ab}\phi_a\psi_b$
of a flavor current, which contain no such loops at this order (section~\ref{sec:oddsusy}).  The odd loops have a small effect on $C_T$, but for $C_J$
they are as large as the $\epsilon^3$ coefficients themselves.

Our results are consistent with emergent supersymmetry. At the $\mathcal N=2$ point of the chiral XY model, superconformal symmetry requires
$C_T/C_{T,\rm free}=C_J^R/C^R_{J,\rm free}$ for the $R$-current.  This identity is not built into our calculation, but we find that it holds up to $\mathcal O(\epsilon^3)$. In $d=3$, two-sided Pad\'e approximants that combine our series with the exact $d=2$
values give $C_T/C_{T,\rm free}=0.714$--$0.730$ at the $\mathcal N=2$ point, around the exact value $0.7268$ from localization. This improves greatly on the result using only $\mathcal O(\epsilon)$
input, which gives $0.77$ \cite{FGKT}.  At the $\mathcal N=1$ point, Pad\'e approximants  give $C_T/C_{T,\rm free}=0.832$--$0.859$, which contains the bootstrap value $0.84207(7)$ \cite{Atanasov}
(table~\ref{tab:d3}).  We also give $C_T$ at the $\mathcal N=1$ fixed points of Liendo and Rong \cite{LiendoRong}, and $C_T$ and $C_J$ at the supersymmetric
fixed points of the matrix Gross--Neveu--Yukawa models~ \cite{PrakashSinha}, including their planar large-$N$ limit
(sections~\ref{sec:LR} and~\ref{sec:matrix}).  At the $SU(3)$ matrix-model point, conjectured to be dual to $\mathcal N=2$ SQED with two flavors
\cite{BeniniBenvenuti}, our $d=3$ estimates contain the values of $C_T$ and $C_J$ obtained by localization \cite{GangYamazaki}.

Using our results, we estimate the universal conductivity at the charge-density-wave, Kekul\'e and N\'eel transitions of graphene,
which correspond to $N=8$ in the chiral Ising, XY and Heisenberg models, to be $\sigma/\sigma_{\rm free}\approx0.91$, $0.79$--$0.83$ and $0.64$--$0.74$
respectively (table~\ref{tab:sigma}). These lie close to the leading large-$N$ values, even at $N=4$.
At the semimetal--superconductor transition on the surface of a topological insulator, where supersymmetry emerges, we show that the exact conductivity
is $\sigma/\sigma_{\rm free}=\frac9{10}\,C_T/C_{T,\rm free}\approx0.654$, $10\%$ below the value reported in \cite{WitczakKrempa}.  The difference arises because the free
theory has two $U(1)$ symmetries, and the electric current is proportional to the superconformal $R$-current only at the interacting fixed point (section~\ref{sec:Rcurrent}).  For the Gross--Neveu--Yukawa model we also give the
$1/N^2$ coefficients $C_{T2}$ and $C_{J2}$ up to $\mathcal O(\epsilon^3)$.  

The paper is organized as follows. In section~\ref{sec:theory} we define the theory and review the two-coupling families of \cite{JOS}, which contain the chiral
Gross--Neveu--Yukawa models.  In section~\ref{sec:method} we describe how we compute $\langle TT\rangle$ and
$\langle JJ\rangle$.  In section~\ref{sec:results} we give $C_T$ and $C_J$ at any fixed point and evaluate them for the two-coupling families, the chiral models
and the Gross--Neveu--Yukawa model at large $N$.  Section~\ref{sec:susy} collects the supersymmetric fixed points and their $d=3$ estimates, and section~\ref{sec:applications}
discusses the Gross--Neveu model in $d=3$ and the universal conductivity.  We conclude in section~\ref{sec:conclusion}.  The appendices contain the Feynman rules (appendix~\ref{app:method}), the checks (appendix~\ref{sec:checks}), details of the loop computation and
the fixed points to $\mathcal O(\epsilon^3)$ (appendix~\ref{app:4loop}), the full $\mathcal O(\epsilon^3)$ result for $C_J$ (appendix~\ref{app:CJ3}), the
odd fermion loops (appendix~\ref{app:odd}), and the supercurrent and flavor-current computation (appendix~\ref{app:supercurrent}).  

\section{Scalar--fermion theories and their fixed points}\label{sec:theory}
\subsection{The general theory}
We consider the general theory of Jack, Osborn and Steudtner (JOS) \cite{JOS} and follow their conventions. It contains $n_s$ real scalars $\phi_a$ and $N$
two-component Majorana fermions $\psi_i$, with real symmetric Yukawa matrices $y^a_{ij}$ that encode the flavor and chirality structure. Dirac fermions are
included by writing each as a pair of Majorana fermions, which makes the $y^a$ block matrices (section~8 of \cite{JOS}). In $d=4-\epsilon$ the action is
\begin{equation}\label{eq:action}
\mathcal{S} =\int d^{4-\eps}x \, \left[\frac12(\partial\phi_a)^2+\bar\psi\,\partial\!\!\!/\,\psi+\phi_a\,\bar\psi\,y^a\psi+\frac{1}{24}\lambda_{abcd}\phi_a\phi_b\phi_c\phi_d\right],
\end{equation}
with $\lambda_{abcd}$ real and symmetric. 

We use Dirac-type propagators and vertices.  A closed fermion loop then gives a spinor trace, a flavor trace and,
for Majorana fermions, a factor of $\frac12$. We write $\Tr$ for the product of all three.  Following \cite{JOS}, the spinor
trace is taken over two-component spinors, so for $N$ two-component Majorana fermions $\Tr\mathbf 1=\tfrac12\cdot 2\cdot N=N$.
Thus $N\equiv\Tr\mathbf 1$ counts fermionic degrees of freedom in units of a single two-component Majorana fermion in $d=3$.
A two-component Dirac fermion in $d=3$ and a four-component Majorana fermion in $d=4$ each have $N=2$, and a four-component
Dirac fermion has $N=4$.  For example, the two valleys and two spin states of graphene give four two-component Dirac fermions, so $N=8$.

After rescaling $y\to4\pi y$ and $\lambda\to16\pi^2\lambda$, the $\overline{\rm MS}$ $\beta$-functions are \cite{JOS}
\begin{equation}\label{eq:rg}
\begin{aligned}
\beta_y^a&=-\tfrac{\epsilon}{2}\,y^a+\tilde\beta_y^a+\gamma_\psi\,y^a+y^a\,\gamma_\psi+\gamma_\phi^{ab}\,y^b,\\
\beta_{\lambda,abcd}&=-\epsilon\,\lambda_{abcd}+\tilde\beta_{\lambda,abcd}+4\,\gamma_\phi^{e(a}\lambda_{bcd)e},
\end{aligned}
\end{equation}
where $\tilde\beta_y$, $\tilde\beta_\lambda$ are the 1PI vertex contributions, $\gamma_\phi^{ab}$ and the flavor matrix $\gamma_\psi$ are the anomalous
dimensions, and parentheses denote symmetrization with unit weight. At one loop, which fixes the normalization,
\begin{equation}\label{eq:rg1}
\gamma_\phi^{ab}=\tfrac12\Tr(y^ay^b),\quad \gamma_\psi=\tfrac12\,y^ay^a,\quad
\tilde\beta_y^a=2\,y^by^ay^b,\quad
\tilde\beta_{\lambda,abcd}=3\,\lambda_{ef(ab}\lambda_{cd)ef}-12\,\Tr\big(y^{(a}y^by^cy^{d)}\big),
\end{equation}
and we use the three-loop results of JOS \cite{JOS}.

\subsection{Two-coupling families}\label{sec:families}
JOS \cite{JOS} observed that a tractable class of fixed points arises by restricting the couplings to
\begin{equation}\label{eq:family}
y^a=y\,t^a\otimes\mathbf 1_m,\qquad \lambda_{abcd}=\lambda\,(\delta_{ab}\delta_{cd}+\delta_{ac}\delta_{bd}+\delta_{ad}\delta_{bc}),
\end{equation}
with $t^a$ ($a=1,\dots,n_s$) traceless $n\times n$ matrices, real symmetric for Majorana and hermitian for Dirac fermions, and $m$ an arbitrary positive integer. This
form has only two independent couplings, $y$ and $\lambda$, and is preserved by the renormalization group if $\mathrm{tr}(t^{(a}t^bt^ct^{d)})\propto\delta^{(ab}\delta^{cd)}$ and
\begin{equation}\label{eq:tid}
t^a t^a=n_s\,\alpha\,\mathbf 1_n,\qquad \mathrm{tr}(t^a t^b)=n\,\alpha\,\delta^{ab},\qquad t^b t^a t^b=\beta\,t^a,\qquad t^b t^c t^a t^b t^c=\gamma\,t^a,
\end{equation}
with the last condition arising from the two-loop beta function. At one loop
\begin{equation}\label{eq:fp1}
y_*^2=\frac{\epsilon}{2K},\qquad K=n_s\alpha+2\beta+\tilde n_f,\qquad \tilde n_f=\tfrac12\alpha N,\qquad
\gamma_\phi=\tilde n_f\,y_*^2,\qquad \gamma_\psi=\tfrac12 n_s\alpha\,y_*^2,
\end{equation}
where $N=\Tr\mathbf 1$ throughout. Table~\ref{tab:families} lists the specific models we focus our attention on here. The chiral $O(k)$
models, with $k$ anticommuting $4\times4$ matrices $\Gamma^a$, are not among the families of JOS \cite{JOS} but have the same form.  The general results of
section~\ref{sec:results} also apply to the other families of \cite{JOS}, including those with a second quartic coupling.
\begin{table}[H]
\centering\footnotesize\setlength{\tabcolsep}{3pt}
\begin{tabular}{lcccccccl}\toprule
family  \cite{JOS} $~~~$ & $t^a$ & fermion & $n$ & $n_s$ & $\alpha$ & $\beta$ & $\gamma$ & model \\ \midrule
diagonal & $\sigma_3$ & Majorana & 2 & 1 & 1 & 1 & 1 & chiral Ising\\
symmetric & $\sigma_3,\sigma_1$ & Majorana & 2 & 2 & 1 & 0 & 4 & chiral XY\\
unitary & $\sigma^a$ & Dirac & 2 & 3 & 1 & $-1$ & 5 & chiral Heisenberg\\
Clifford & $\Gamma^a$ & Dirac & 4 & $k$ & 1 & $2-k$ & $k-(k-1)(k-4)$ & chiral $O(k)$\\ \bottomrule
\end{tabular}
\caption{Some scalar-fermion theories belonging to the two-coupling families \eqref{eq:family} of JOS.}
\label{tab:families}
\end{table}

\section{Computing \texorpdfstring{$\langle TT\rangle$ and $\langle JJ\rangle$}{<TT> and <JJ>}}\label{sec:method}

A continuous flavor symmetry acts as $\delta\psi=i\Theta\psi$, $\delta\phi=\theta\,\phi$, with $\theta$ real antisymmetric and $\Theta$ hermitian.  For Majorana fields $i\Theta$ must be real, so $\Theta$ is imaginary and antisymmetric. The action \eqref{eq:action} is invariant if and only if,
\begin{equation}\label{eq:ward}
\begin{gathered}
i\,[y^a,\Theta]=\theta_{ab}\,y^b,\\
\theta_{ae}\lambda_{ebcd}+\theta_{be}\lambda_{aecd}+\theta_{ce}\lambda_{abed}+\theta_{de}\lambda_{abce}=0 .
\end{gathered}
\end{equation}

The conserved current, which we take to be the Noether current divided by $i$, and the stress tensor are
\begin{equation}\label{eq:TJ}
\begin{aligned}
J_\mu&=\bar\psi\,\gamma_\mu \Theta\psi-i\,(\partial_\mu\phi_a)\,\theta_{ab}\,\phi_b,\\
T_{\mu\nu}&=\tfrac12\,\bar\psi\,\gamma_{(\mu}\overleftrightarrow{\partial}_{\!\nu)}\psi+\partial_\mu\phi_a\,\partial_\nu\phi_a
-\xi\,\big(\partial_\mu\partial_\nu-\delta_{\mu\nu}\partial^2\big)(\phi_a\phi_a)-\delta_{\mu\nu}\,\mathcal L,
\end{aligned}
\end{equation}
where $\overleftrightarrow\partial=\overrightarrow\partial-\overleftarrow\partial$, $\mathcal L$ is the Lagrangian density of \eqref{eq:action}, and
$\xi=(d-2)/\big(4(d-1)\big)$ ensures the free scalar stress tensor is traceless.

We normalize $C_J$ by the Noether current $iJ$, so that $C_J>0$ as in \cite{OsbornPetkou,Diab}. For the current $J$ used in our computation this means
$\langle J_\mu(x)J_\nu(0)\rangle=-C_J\,I_{\mu\nu}(x)/x^{2(d-1)}$. Our results are ratios $C_J/C_{J,\rm free}$, which do not depend on this sign.

\subsection{Operators and projections}\label{sec:ops}

In momentum space, conservation of $T_{\mu\nu}$ and $J_\mu$ makes the two-point functions transverse up to contact terms (these involve one-point functions, which vanish for massless fields in dimensional regularization). Rotational invariance and transversality therefore restrict the correlators to the form
\begin{equation}\label{eq:decomp}
\begin{aligned}
\langle T_{\mu\nu}(Q)\,T_{\rho\sigma}(-Q)\rangle&=Q^4\big[P_{\mu\nu,\rho\sigma}(Q)\,A_T(Q^2)+\pi_{\mu\nu}\pi_{\rho\sigma}\,B_T(Q^2)\big],\\
\langle J_\mu(Q)\,J_\nu(-Q)\rangle&=Q^2\,\pi_{\mu\nu}\,A_J(Q^2),
\end{aligned}
\end{equation}
where $\pi_{\mu\nu}$ and $P_{\mu\nu,\rho\sigma}$ are the usual transverse projector and the transverse-traceless spin-two projector,
\begin{equation}
\begin{aligned}\label{eq:projector}
    \pi_{\mu\nu} &=\delta_{\mu\nu}-Q_\mu Q_\nu/Q^2,\\
    P_{\mu\nu,\rho\sigma} &=\tfrac12\big(\pi_{\mu\rho}\pi_{\nu\sigma}+\pi_{\mu\sigma}\pi_{\nu\rho}\big)-\frac{1}{d-1}\pi_{\mu\nu}\pi_{\rho\sigma}.
\end{aligned}
\end{equation} 

At the fixed point the trace $T^\mu{}_\mu$ vanishes, so the spin-zero part $B_T$ is a contact term, and projecting onto spin two eliminates it.   We isolate $A_T$ and $A_J$ by contracting the stress-tensor and current two-point functions with the transverse-traceless and transverse projectors, respectively:
\begin{equation}\label{eq:ATAJ}
\begin{aligned}
A_T(Q^2)&=\frac{2P_{\mu\nu,\rho\sigma}(Q)\,
\langle T_{\mu\nu}(Q)T_{\rho\sigma}(-Q)\rangle}{(d+1)(d-2)(Q^2)^2},\\
A_J(Q^2)&=\frac{\pi_{\mu\nu}(Q)\,
\langle J_\mu(Q)J_\nu(-Q)\rangle}{(d-1)Q^2}.
\end{aligned}
\end{equation}

At a fixed point the position-space correlators take the conformal form, and their Fourier transform gives, up to contact terms,
\begin{equation}\label{eq:NTNJ}
A_T=N_T\,C_T\,(Q^2)^{-\epsilon/2},\qquad A_J=N_J\,C_J\,(Q^2)^{-\epsilon/2},
\end{equation}
with
\begin{equation*}
N_T=\frac{d-1}{d-2}\,\frac{\pi^{d/2}\,\Gamma\big(2-\frac d2\big)}{2^{d-2}\,\Gamma(d+2)},\qquad
N_J=\frac{\pi^{d/2}\,\Gamma\big(2-\frac d2\big)}{2^{d-3}\,\Gamma(d)}.
\end{equation*}

Each $L$-loop diagram, computed with bare couplings, contributes a pure power $(Q^2)^{-L\epsilon/2}$ to $A_T$ or $A_J$.  Only after the couplings are renormalized and set to their fixed-point values do these powers combine into the single power of \eqref{eq:NTNJ}.  We therefore Fourier transform each power separately to position space, together with its tensor factor in \eqref{eq:decomp}, and read off $C_T$ and $C_J$ there.  Working at $x\neq0$ has a further advantage: terms polynomial in $Q$, such as the $1/\epsilon$ pole of the one-loop $A_T$, are contact terms and drop out, so only the couplings need renormalization (section~\ref{sec:renorm}).  Appendix~\ref{app:method} gives the Feynman rules and the normalizations of the projection and the Fourier transform.

At one loop the diagrams give the free values, which are exact for all $d$,
\begin{equation}\label{eq:free}
C_{T,\rm free}=\left[n_s+\frac{(d-1)\,\Tr\mathbf 1}{2}\right]\,C_{T,S},\qquad
C_{J,\rm free}=\left[\Tr(\Theta \Theta)+\frac{\theta_{ab}\theta_{ab}}{d-2}\right]\,C_{J,\psi},
\end{equation}
where $C_{T,S}=d/\big((d-1)S_d^2\big)$ is the value for a free real scalar, $C_{J,\psi}=1/S_d^2$ is the value for free fermions with $\Tr(\Theta\Theta)=1$, and
$S_d=2\pi^{d/2}/\Gamma(d/2)$.  We call the two terms of $C_{T,\rm free}$ its scalar and fermion parts, $C_{T,\sigma}$ and $C_{T,\rm ferm}$.  Ratios
$C_T/C_{T,\rm free}$ and $C_J/C_{J,\rm free}$ are always taken with respect to the free values for the same field content and the same current, except
for the $R$-current, whose free value uses the $R$-charges of the free theory (section~\ref{sec:Rcurrent}).

\subsection{Diagrams up to three loops}\label{sec:diagrams}
We generate all Wick contractions of $\langle O(x)O(0)\rangle$, with $O=T$ or $J$, using $n_Y$ Yukawa and $n_Q$ quartic vertices with
$n_Y/2+n_Q\le2$. 

\begin{figure}[htbp]\centering
\begin{minipage}[b]{0.30\textwidth}\centering\resizebox{\linewidth}{!}{\begin{tikzpicture}[scale=0.78]
  \useasboundingbox (-3.0,-2.75) rectangle (3.0,2.75);
  \draw[scalar] (-0.7,1.212) to (-0.4,2.3);
  \draw[scalar] (0.7,1.212) to (0.4,2.3);
  \draw[fermion] (-1.4,0) to[bend left=30] (-0.7,1.212);
  \draw[fermion] (-0.7,1.212) to[bend left=30] (0.7,1.212);
  \draw[fermion] (0.7,1.212) to[bend left=30] (1.4,0);
  \draw[fermion] (1.4,0) to[out=-90,in=-90,looseness=1.7] (-1.4,0);
  \draw[fermion] (-0.4,2.3) to[bend left=70] (0.4,2.3);
  \draw[fermion] (0.4,2.3) to[bend left=70] (-0.4,2.3);
  \fill (-0.7,1.212) circle (0.055);
  \fill (0.7,1.212) circle (0.055);
  \fill (-0.4,2.3) circle (0.055);
  \fill (0.4,2.3) circle (0.055);
  \Tpsi{-1.4}{0}
  \Tpsi{1.4}{0}
\end{tikzpicture}}\\[1pt]{\scriptsize A1}\end{minipage}\hfill\begin{minipage}[b]{0.30\textwidth}\centering\resizebox{\linewidth}{!}{\begin{tikzpicture}[scale=0.78]
  \useasboundingbox (-3.0,-2.75) rectangle (3.0,2.75);
  \draw[scalar] (0,1.4) to (0,0.4);
  \draw[scalar] (0,-1.4) to (0,-0.4);
  \draw[fermion] (-1.4,0) to[bend left=45] (0,1.4);
  \draw[fermion] (0,1.4) to[bend left=45] (1.4,0);
  \draw[fermion] (1.4,0) to[bend left=45] (0,-1.4);
  \draw[fermion] (0,-1.4) to[bend left=45] (-1.4,0);
  \draw[fermion] (0,0.4) to[bend left=70] (0,-0.4);
  \draw[fermion] (0,-0.4) to[bend left=70] (0,0.4);
  \fill (0,1.4) circle (0.055);
  \fill (0,-1.4) circle (0.055);
  \fill (0,0.4) circle (0.055);
  \fill (0,-0.4) circle (0.055);
  \Tpsi{-1.4}{0}
  \Tpsi{1.4}{0}
\end{tikzpicture}}\\[-1pt]{\scriptsize A3}\end{minipage}\hfill\begin{minipage}[b]{0.30\textwidth}\centering\resizebox{\linewidth}{!}{\begin{tikzpicture}[scale=0.92]
  \useasboundingbox (-3.45,-1.35) rectangle (3.45,1.35);
  \draw[scalar] (-0.85,0.9) to (0.15,0.78);
  \draw[scalar] (0.95,0.56) to (2.1,0);
  \draw[scalar] (-0.85,-0.9) to (2.1,0);
  \draw[fermion] (-2.1,0) to[bend left=20] (-0.85,0.9);
  \draw[fermion] (-0.85,0.9) to[bend left=20] (-0.85,-0.9);
  \draw[fermion] (-0.85,-0.9) to[bend left=20] (-2.1,0);
  \draw[fermion] (0.15,0.78) to[bend left=70] (0.95,0.56);
  \draw[fermion] (0.95,0.56) to[bend left=70] (0.15,0.78);
  \fill (-0.85,0.9) circle (0.055);
  \fill (-0.85,-0.9) circle (0.055);
  \fill (0.15,0.78) circle (0.055);
  \fill (0.95,0.56) circle (0.055);
  \Tpsi{-2.1}{0}
  \Tsig{2.1}{0}
\end{tikzpicture}}\\[-1pt]{\scriptsize B1}\end{minipage}\\[6pt]
\begin{minipage}[b]{0.30\textwidth}\centering\resizebox{\linewidth}{!}{\begin{tikzpicture}[scale=0.92]
  \useasboundingbox (-3.45,-1.45) rectangle (3.45,1.45);
  \draw[scalar] (-2.0,0) to (-1.3,0.85);
  \draw[scalar] (-0.45,1.05) to (0.45,1.05);
  \draw[scalar] (1.3,0.85) to (2.0,0);
  \draw[scalar] (-2.0,0) to[out=-45,in=-135,looseness=0.9] (2.0,0);
  \draw[fermion] (-0.45,1.05) to[bend left=70] (-1.3,0.85);
  \draw[fermion] (-1.3,0.85) to[bend left=70] (-0.45,1.05);
  \draw[fermion] (0.45,1.05) to[bend left=70] (1.3,0.85);
  \draw[fermion] (1.3,0.85) to[bend left=70] (0.45,1.05);
  \fill (-1.3,0.85) circle (0.055);
  \fill (-0.45,1.05) circle (0.055);
  \fill (0.45,1.05) circle (0.055);
  \fill (1.3,0.85) circle (0.055);
  \Tsig{-2.0}{0}
  \Tsig{2.0}{0}
\end{tikzpicture}}\\[-1pt]{\scriptsize C1}\end{minipage}\hfill\begin{minipage}[b]{0.30\textwidth}\centering\resizebox{\linewidth}{!}{\begin{tikzpicture}[scale=0.92]
  \useasboundingbox (-3.45,-1.45) rectangle (3.45,1.45);
  \draw[scalar] (-2.0,0) to (-0.42,0.95);
  \draw[scalar] (0.42,0.95) to (2.0,0);
  \draw[scalar] (-2.0,0) to (-0.42,-0.95);
  \draw[scalar] (0.42,-0.95) to (2.0,0);
  \draw[fermion] (0.42,0.95) to[bend left=70] (-0.42,0.95);
  \draw[fermion] (-0.42,0.95) to[bend left=70] (0.42,0.95);
  \draw[fermion] (0.42,-0.95) to[bend left=70] (-0.42,-0.95);
  \draw[fermion] (-0.42,-0.95) to[bend left=70] (0.42,-0.95);
  \fill (-0.42,0.95) circle (0.055);
  \fill (0.42,0.95) circle (0.055);
  \fill (-0.42,-0.95) circle (0.055);
  \fill (0.42,-0.95) circle (0.055);
  \Tsig{-2.0}{0}
  \Tsig{2.0}{0}
\end{tikzpicture}}\\[-1pt]{\scriptsize C2}\end{minipage}\hfill\begin{minipage}[b]{0.30\textwidth}\centering\resizebox{\linewidth}{!}{\begin{tikzpicture}[scale=0.92]
  \useasboundingbox (-3.45,-1.3) rectangle (3.45,1.3);
  \draw[scalar] (-0.85,0.9) to (0.85,0.9);
  \draw[scalar] (-0.85,-0.9) to (0.85,-0.9);
  \draw[fermion] (-2.1,0) to[bend left=20] (-0.85,0.9);
  \draw[fermion] (-0.85,0.9) to[bend left=20] (-0.85,-0.9);
  \draw[fermion] (-0.85,-0.9) to[bend left=20] (-2.1,0);
  \draw[fermion] (2.1,0) to[bend right=20] (0.85,0.9);
  \draw[fermion] (0.85,0.9) to[bend right=20] (0.85,-0.9);
  \draw[fermion] (0.85,-0.9) to[bend right=20] (2.1,0);
  \fill (-0.85,0.9) circle (0.055);
  \fill (-0.85,-0.9) circle (0.055);
  \fill (0.85,0.9) circle (0.055);
  \fill (0.85,-0.9) circle (0.055);
  \Tpsi{-2.1}{0}
  \Tpsi{2.1}{0}
\end{tikzpicture}}\\[-1pt]{\scriptsize D1}\end{minipage}
\caption{The three-loop $\langle TT\rangle$ classes with two fermion loops and flavor invariant $\Tr(y^ay^b)\Tr(y^ay^b)$, which are leading at large $N$.
There are eleven classes, with six distinct values.  The other five (A2, B2--B4, D2) follow from A1, B1 and D1 by reversing a fermion loop or
exchanging $x\leftrightarrow0$, which leaves the value unchanged.  A: dressed scalar exchange inside a $T_\psi$ loop; B: fermion triangle with
both scalar lines on $T_\phi$; C: scalar loop with two fermion bubbles; D: the Aslamazov--Larkin diagram, the only one that is not a bubble insertion.
Solid lines: fermions; dashed: scalars; circles: $T_\psi$; squares: $T_\phi$; $T(x)$ on the left, $T(0)$ on the right.}
\label{fig:3loop}
\end{figure}
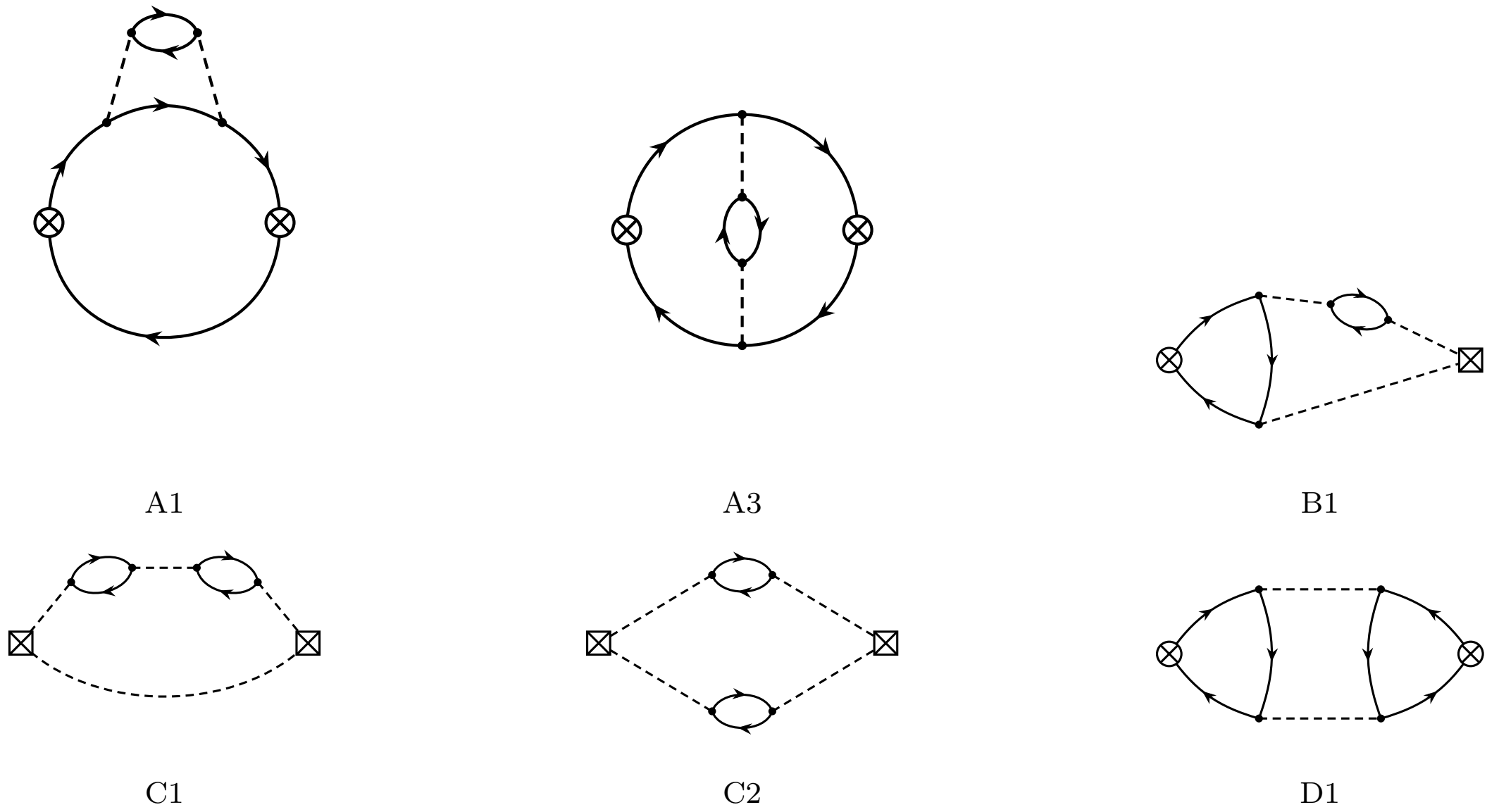

To keep track of combinatorial factors, we group the Wick contractions into isomorphism classes. Each class is essentially a Feynman diagram. Two Wick contractions belong to the same class if relabeling the internal vertices maps one onto the
other, with $x$ and $0$, the operator component ($T_\psi$ or $T_\phi$, $J_\psi$ or $J_\phi$) at each, and the orientation of every fermion
line held fixed.  The weight of a class is the number of its contractions divided by $n_Y!\,n_Q!\,(4!)^{n_Q}$, the factor from expanding
$e^{-S_{\rm int}}$, which is the inverse of the usual symmetry factor. Because the endpoints and fermion orientations are fixed, a diagram and its image under $x\leftrightarrow0$ or under the reversal of a fermion loop count as separate classes, such as A1 and A2 (figure~\ref{fig:3loop}).    

Three kinds of graph vanish and are dropped:
\begin{itemize}
\item disconnected graphs, which contain one-point functions;
\item graphs in which a single vertex separates $x$ from $0$: each side depends only on $Q$, so it is proportional to $Q_\mu Q_\nu$ or
$\delta_{\mu\nu}$ (or $Q_\mu$ for a current), and the projection removes it;
\item graphs with a self-loop, or with any part attached to the rest through a single vertex, which are scaleless.
\end{itemize}
In particular, no $y^2\lambda$ diagram contributes, since in each the quartic vertex either separates $x$ from $0$ or carries a self-loop.

Each class comes with a coupling-tensor invariant that can be read off directly from its graph: fermion loops become traces, taken in the order of the fermion
arrows, and scalar lines become index contractions.  We expand these invariants on the bases of \eqref{eq:CT} and \eqref{eq:CJ} by evaluating
them numerically, and reconstruct the rational coefficients.  For $\langle TT\rangle$ we evaluate them on random tensors.  For $\langle JJ\rangle$
we use random theories with an exact $U(1)$ or $SU(2)$ symmetry, so that the Ward identity \eqref{eq:ward} holds.  The Ward identity turns any
$\theta$ contracted into a Yukawa coupling into a commutator with $\Theta$, so the basis needs $\theta$ only where it is contracted with $\lambda$.

The four-loop diagrams are generated in the same way (see appendix~\ref{app:4loop}). Table~\ref{tab:classes} in appendix~\ref{app:4loop} gives the number of classes at each order.  Figure~\ref{fig:3loop} shows the double-trace classes of $\langle TT\rangle$,
which dominate at large $N$.  The accompanying repository~\cite{yukawactcj}  provides the code used to enumerate the classes and lists every class with its graph, weight, value and flavor decomposition.

\subsection{Integrals, regularization and renormalization}\label{sec:renorm}

An $L$-loop term in momentum space is a pure power $(Q^2)^{-L\epsilon/2}$, and its Fourier transform (appendix~\ref{app:method}) carries a factor
$1/\Gamma(L\epsilon/2)\approx L\epsilon/2$.  A $1/\epsilon$ pole in momentum space therefore contributes at $\mathcal O(\epsilon^0)$ in position space.
Since the couplings of an $L$-loop diagram are $\mathcal O(\epsilon^{L-1})$ at the fixed point, the diagram contributes to $\mathcal O(\epsilon^k)$
through the $\epsilon^{k-L}$ term of its momentum-space value.  For the $\mathcal O(\epsilon^2)$ results we need to evaluate the three-loop integrals up to order
$1/\epsilon$ and the two-loop integrals up to order $\epsilon^0$; for $\mathcal O(\epsilon^3)$, we need the four-loop integrals up to order $1/\epsilon$, the three-loop
integrals up to order $\epsilon^0$ and the two-loop integrals up to order $\epsilon^1$.

After the projection of section~\ref{sec:ops}, every diagram is an integral involving  massless scalar propagators.  We take the Dirac traces in FORM
\cite{FORM,FORM4,FORM5} and reduce the integrals by integration by parts with Forcer \cite{Forcer}.  Forcer works in $D=4-2\epsilon'$, so
$\epsilon'=\epsilon/2$, and we normalize each $L$-loop result in the G-scheme, $X=\text{integral}/(\epsilon'G(1,1))^L$ at $Q^2=1$
(eq.~\eqref{eq:app-forcer-output}).\footnote{Forcer expands each integral to a chosen depth in $\epsilon'$, and the highest orders of a finite expansion
can be incomplete.  We expand to depth 5 for diagrams up to three loops and to depth 3 at four loops.  Recomputing 73 four-loop classes (7 for
$\langle TT\rangle$, 66 for $\langle JJ\rangle$) at depth 4 leaves every coefficient through $\epsilon'^{-1}$ unchanged.}

Since $T$ and $J$ are conserved, their projected parts need no renormalization of their own, and only the couplings renormalize.  We use minimal
subtraction.  The G-scheme and $\overline{\rm MS}$ differ from it only by loop-dependent factors that leave the RG functions unchanged \cite{CKT,Forcer},
so our counterterms are consistent with the $\overline{\rm MS}$ $\beta$-functions of \cite{JOS}.  Up to three loops it suffices to take
$y_0=\mu^{\epsilon/2}\big(y+\beta_1(y)/\epsilon\big)$, where $\beta_1$ is the $\epsilon$-independent one-loop part of $\beta_y$ in \eqref{eq:rg}.  The
quartic coupling first enters through $\lambda^2$ at three loops and needs no counterterm at this order; the four-loop counterterms are given in
appendix~\ref{app:4loop}.  We express the bare series in renormalized couplings and $\log\mu^2x^2$, expanding in the couplings first and in
$\epsilon$ second, and then insert the fixed point.  The poles cancel separately in each flavor invariant, and the remaining dependence on
$\log\mu^2x^2$ multiplies only terms proportional to $\beta_y$, so it drops out at a fixed point.

\subsection{Dirac algebra and continuation to $d=3$}\label{sec:dirac}

We are mainly interested in continuing our results to $d=3$.  As the literature~\cite{Zerf,ChakrabortyMoch} has observed, this raises subtleties at higher orders in
perturbation theory, because $\gamma$-matrices in four and three dimensions differ.  We distinguish two classes of theories:
\begin{itemize}
    \item theories that have a natural four-dimensional counterpart, whose two-component three-dimensional spinors pair into four-component
    four-dimensional spinors;
    \item theories with no four-dimensional counterpart, such as one with an odd number of two-component Majorana spinors.  Every theory with only
    $\mathcal N=1$ supersymmetry in $d=3$ belongs to this class.
\end{itemize}

We use $\{\gamma_\mu,\gamma_\nu\}=2\delta_{\mu\nu}\mathbf 1$ with $\delta_{\mu\mu}=D$, which need not be an integer, and keep $\mathrm{tr}\,\mathbf 1$ as a $D$-independent
factor, which we absorb into $N=\Tr\mathbf 1$, as in \cite{Diab,Zerf}.  For theories with a four-dimensional counterpart, we set traces of an odd
number of $\gamma$-matrices to zero, as in four dimensions, so fermion loops with an odd number of Yukawa vertices vanish.  The three-loop
$\beta$-functions of \cite{JOS} use the same conventions.  Since the flavor matrices carry chirality, no $\gamma_5$ appears, and its ambiguities
\cite{tHooftVeltman,Jegerlehner,ChetyrkinZoller,PooleThomsen} do not arise.\footnote{A formulation with an
anticommuting $\gamma_5$ in the Yukawa vertices could differ only through parity-odd traces, and these cannot contribute at four loops: after
projection the correlators depend on $Q$ alone, so a single $\epsilon$-tensor cannot survive and two parity-odd fermion loops are needed.  Each
needs three independent loop momenta, hence at least four Yukawa vertices, so at least eight in total, i.e., five loops \cite{ChakrabortyMoch}.}

For theories with no four-dimensional counterpart, we need to modify this prescription.  In non-integral dimensions, the
$\gamma_a$, $a=1,\dots,D$, are not matrices of any fixed size.  We treat them as symbols that obey $\{\gamma_a,\gamma_b\}=2\delta_{ab}$ with
$\delta_{aa}=D$, and define formal rules for evaluating their trace.  For an even number of $\gamma$'s, we use the anticommutator to move the symbols past each other
until only $\mathrm{tr}\,\mathbf 1$ remains; the result, a polynomial in Kronecker deltas, is cyclic.  For an odd number, we need an additional
rule.  Following Zerf \textit{et al.} \cite{Zerf}, who computed the four-loop $\beta$-functions, we continue theories with out a four-dimensional counterpart using
the three-dimensional rule
\begin{equation}\label{eq:tr3}
\mathrm{tr}(\gamma_a\gamma_b\gamma_c)=i\,\tilde\epsilon_{abc}\,\mathrm{tr}\,\mathbf 1 .
\end{equation}
Here $\tilde\epsilon_{abc}$ continues the three-dimensional Levi-Civita symbol to $D$ dimensions.  It is totally antisymmetric, and its product is formally defined to be the determinant of Kronecker deltas,
\begin{equation}\label{eq:epseps}
\tilde\epsilon_{abc}\,\tilde\epsilon_{def}=\det\begin{pmatrix}\delta_{ad}&\delta_{ae}&\delta_{af}\\ \delta_{bd}&\delta_{be}&\delta_{bf}\\
\delta_{cd}&\delta_{ce}&\delta_{cf}\end{pmatrix},\qquad\text{so that, e.g.,}\qquad \tilde\epsilon_{abc}\tilde\epsilon_{abc}=D(D-1)(D-2).
\end{equation}
In $D=3$, $\tilde\epsilon$ is the familiar Levi-Civita symbol, but for $D\neq3$ no actual tensor satisfies these relations.  Like the $\gamma$-matrices, we treat $\tilde\epsilon$ as a formal symbol in $D$ dimensions, defined to obey the rules \eqref{eq:tr3} and
\eqref{eq:epseps}.  (We remark that these rules can be formulated precisely in the language of Deligne categories, in which $D$ is a continuous parameter; we
will present this formulation, and its implications for the odd fermion loops, in a separate paper.)  

In particular, at $D=4$ these rules do not describe $4\times4$ $\gamma$-matrices, whose odd traces vanish: they define a
continuation of the three-dimensional theory, which is all that the $\epsilon$-expansion of a theory without a four-dimensional counterpart
requires.  With these rules, fermion loops with an odd number of Yukawa vertices survive.  In the $\beta$-functions they first appear at four loops, where other authors
\cite{Zerf,ChakrabortyMoch} found that, using \eqref{eq:tr3} and \eqref{eq:epseps}, they restore supersymmetry at the $\mathcal N=1$ point.

For our computation, however, this prescription alone does not remove all the ambiguities.  In $\langle TT\rangle$ and $\langle JJ\rangle$ it first
affects four-loop diagrams with two fermion loops, each containing the operator and three Yukawa vertices (figure~\ref{fig:odd}). Unlike the diagrams of \cite{Zerf}, which involve traces of three $\gamma$-matrices,
these contain traces of five $\gamma$-matrices as shown in figure~\ref{fig:open}.

We evaluate a trace of five $\gamma$'s as follows. We first use the anticommutator to reduce the product of five $\gamma$'s to antisymmetrized products of five, three and one $\gamma$, with Kronecker deltas as coefficients.  For $\gamma$-matrices in three dimensions, only the anti-symmetrized product of three $\gamma$'s has a non-vanishing trace; we promote this to a formal rule governing traces of antisymmetrized products of any odd number of gamma matrices. Then,
\begin{equation}\label{eq:five}
\mathrm{tr}(\gamma_{a_1}\gamma_{a_2}\gamma_{a_3}\gamma_{a_4}\gamma_{a_5})=i\,\mathrm{tr}\,\mathbf 1\sum_{i<j}(-1)^{j-i-1}\,\delta_{a_ia_j}\,
\tilde\epsilon_{\{a\}\setminus\{a_i,a_j\}} ,
\end{equation}
where $\{a\}\setminus\{a_i,a_j\}$ denotes the remaining three indices in their original order.  This expression depends on the order in which we
write the five symbols, and in particular on which one comes first.  The trace of finite-dimensional matrices would be cyclic, but the $\gamma_a$ are not finite-dimensional
matrices, and \eqref{eq:five} is not cyclic: starting the product one symbol later changes it by
\begin{equation}\label{eq:cyc}
\mathrm{tr}(\gamma_{1}\gamma_{2}\gamma_{3}\gamma_{4}\gamma_{5})-\mathrm{tr}(\gamma_{2}\gamma_{3}\gamma_{4}\gamma_{5}\gamma_{1})
=2i\,\mathrm{tr}\,\mathbf 1\,\big(\delta_{12}\tilde\epsilon_{345}-\delta_{13}\tilde\epsilon_{245}+\delta_{14}\tilde\epsilon_{235}-\delta_{15}\tilde\epsilon_{234}\big),
\end{equation}
where $\gamma_i\equiv\gamma_{a_i}$ and $\delta_{ij}\equiv\delta_{a_ia_j}$.  We say that we \emph{open} the trace at the symbol we write first.  For
the loop of figure~\ref{fig:open}, opening it at the operator gives $\mathrm{tr}(z\!\!\!/\,\,p\!\!\!/_1\,p\!\!\!/_2\,p\!\!\!/_3\,p\!\!\!/_4)$, and
opening it at the Yukawa vertex $b$ gives $\mathrm{tr}(p\!\!\!/_3\,p\!\!\!/_4\,z\!\!\!/\,\,p\!\!\!/_1\,p\!\!\!/_2)$.  In $d=4$ both vanish, and in
$d=3$ they agree.

The bracket in \eqref{eq:cyc} is antisymmetric in the four indices $a_2,\dots,a_5$.  In three dimensions it vanishes, because four indices that
take three values cannot all differ (the Schouten identity).  In $D$ dimensions it does not.  Any complete contraction of it, which we compute using
\eqref{eq:epseps}, is a polynomial in $D$ that vanishes at $D=3$, and so carries a factor $D-3$.  In the diagrams, the other fermion loop supplies
a second $\tilde\epsilon$, and contracting the two using \eqref{eq:epseps} gives, for example,
\begin{equation}
\big(\delta_{12}\tilde\epsilon_{345}-\delta_{13}\tilde\epsilon_{245}+\delta_{14}\tilde\epsilon_{235}-\delta_{15}\tilde\epsilon_{234}\big)\,
\delta_{12}\tilde\epsilon_{345}=(D-3)\,D(D-1)(D-2),
\end{equation}
which is not small near $d=4$.  We computed these diagrams for every choice, and their coefficients depend, by factors of order one, on where we
open the traces.

\begin{figure}[htbp]\centering
\begin{tikzpicture}[scale=1.15]
  \draw[fermion] (-1.1,0) arc (180:90:1.1);
  \draw[fermion] (0,1.1) arc (90:0:1.1);
  \draw[fermion] (1.1,0) arc (0:-90:1.1);
  \draw[fermion] (0,-1.1) arc (-90:-180:1.1);
  \draw[scalar] (0,1.1) -- (0,1.75); \draw[scalar] (1.1,0) -- (1.75,0); \draw[scalar] (0,-1.1) -- (0,-1.75);
  \foreach \p in {(0,1.1),(1.1,0),(0,-1.1)} \fill \p circle (0.055);
  \node[font=\small,above left] at (0,1.15) {$a$}; \node[font=\small,below right] at (1.15,0) {$b$}; \node[font=\small,below left] at (0,-1.15) {$c$};
  \node[font=\small] at (135:1.45) {$p_1$}; \node[font=\small] at (45:1.45) {$p_2$}; \node[font=\small] at (-45:1.45) {$p_3$}; \node[font=\small] at (-135:1.45) {$p_4$};
  \Tpsi{-1.1}{0}
  \node[font=\small,left=6pt] at (-1.1,0) {$z\!\!\!/$};
  \draw[BrickRed,very thick] (-1.1,0) circle (0.32);
  \draw[MidnightBlue,very thick] (1.1,0) circle (0.22);
  \node[anchor=west,align=left,font=\small] at (2.4,0.35) {\textcolor{BrickRed}{opened at the operator:}\quad $\mathrm{tr}(z\!\!\!/\,\,p\!\!\!/_1\,p\!\!\!/_2\,p\!\!\!/_3\,p\!\!\!/_4)$};
  \node[anchor=west,align=left,font=\small] at (2.4,-0.35) {\textcolor{MidnightBlue}{opened at $b$:}\quad $\mathrm{tr}(p\!\!\!/_3\,p\!\!\!/_4\,z\!\!\!/\,\,p\!\!\!/_1\,p\!\!\!/_2)$};
\end{tikzpicture}
\caption{One fermion loop of the four-loop diagrams of figure~\ref{fig:odd}: the operator ($\otimes$, contributing $z\!\!\!/$) and three Yukawa
vertices $a,b,c$, joined by four propagators.  Its trace contains five $\gamma$-matrices.  It can be opened at any of them; two choices are
shown, which agree in $d=3$ but differ in $D$ dimensions by terms of the form \eqref{eq:cyc}.}
\label{fig:open}
\end{figure}

To remove the ambiguity, we use $\mathcal N=1$ supersymmetry, which relates the two-point functions of the spin-$\frac32$ supercurrent $S$ and of the
spin-$\frac12$ fermionic partner $j$ of a flavor current to those of $T$ and $J$.  These two-point functions have no odd fermion loops at this order,
so computing them fixes the continuation (section~\ref{sec:oddsusy}): both traces are opened at the two ends of the same scalar line, and the two
$\tilde\epsilon$-tensors are contracted with \eqref{eq:epseps}, as in \cite{Zerf}.  The result does not depend on which of the three scalar lines
we use, and it respects the Ward identity of the current.  For three-dimensional-only theories without supersymmetry we use the same continuation,
which amounts to making it part of the regularization scheme.  The resulting terms, \eqref{eq:oddT} and \eqref{eq:oddJ}, are the last terms of
\eqref{eq:CT3} and \eqref{eq:CJ3}; appendix~\ref{app:odd} gives the details.

These terms vanish identically for theories obtained from four dimensions \cite{JOS}, and they must be evaluated with the actual three-dimensional
flavor structure of the theory.  For example, a four-dimensional Majorana fermion gives two two-component Majorana fermions with opposite signs of
$y$, and $\Tr(y^ay^by^c)=0$; the same two fermions with equal signs form a two-component Dirac fermion with a $U(1)$ symmetry, a different
three-dimensional theory with $\Tr(y^ay^by^c)\neq0$.  A representation that reproduces only the even invariants, such as $y\propto\mathbf 1$ for the
Gross--Neveu--Yukawa model, cannot give these terms.

The checks of the computation are collected in appendix~\ref{sec:checks}.

\section{Results}\label{sec:results}
In this section we give $C_T$ and $C_J$ at an arbitrary fixed point of \eqref{eq:action} (section~\ref{sec:res2}), reduce them for the
two-coupling families of section~\ref{sec:families} (section~\ref{sec:famres}), evaluate them for the chiral Ising, XY and Heisenberg models
(section~\ref{sec:models}), and expand the Gross--Neveu--Yukawa results in $1/N$ (section~\ref{sec:largeN}).  The supersymmetric fixed points are
treated separately in section~\ref{sec:susy}.

\subsection{\texorpdfstring{$C_T$ and $C_J$ up to $\mathcal O(\epsilon^3)$}{C\_T and C\_J up to O(epsilon\^{}3)}}\label{sec:res2}

We write traces of products of Yukawa matrices as words, $\Tr(ab\cdots)\equiv\Tr(y^ay^b\cdots)$, in which the letter $\Theta$ stands for the
generator of the current; for example, $\Tr(\Theta\Theta ab)=\Tr(\Theta\Theta y^ay^b)$.  Repeated indices are summed, and
$\lambda\cdot\lambda\equiv\lambda_{abcd}\lambda_{abcd}$.

At a fixed point the couplings are series in $\epsilon$, $y_*^a=\epsilon^{1/2}(Y_1^a+\epsilon Y_2^a+\dots)$ and $\lambda_*=\epsilon(L_1+\epsilon L_2+\dots)$.
An invariant with $n_y$ Yukawa and $n_\lambda$ quartic couplings is therefore of order $\epsilon^{n_y/2+n_\lambda}$, and the invariants from $L$-loop
diagrams have $n_y/2+n_\lambda=L-1$.  The free terms together with the two- and three-loop invariants therefore give $C_T$ and $C_J$ through
$\mathcal O(\epsilon^2)$.  At $\mathcal O(\epsilon^3)$ two kinds of term contribute: the four-loop invariants, at their leading order, and the two- and
three-loop invariants, through the higher-order terms $Y_2$, $Y_3$ and $L_2$ of the fixed-point couplings, which require the three-loop $\beta_y$
(appendix~\ref{app:fp3}).

\paragraph{Stress tensor.}
At any fixed point,
\begin{empheq}[box=\fbox]{align}
\frac{C_T}{C_{T,S}}={}&n_s+\tfrac12(d-1)\,\Tr\mathbf 1-\tfrac54\,\Tr(aa)-\tfrac16\,\Tr(ab)\Tr(ab)\nonumber\\
&-\tfrac{11}{48}\,\Tr(aabb)-\tfrac{19}{6}\,\Tr(abab)-\tfrac{5}{36}\,\lambda_{abcd}\lambda_{abcd}\label{eq:CT}\\
&+\big(\tfrac{1}{72}+\tfrac98\zeta_3\big)\Tr(aabbcc)+\big(\tfrac{21}{2}\zeta_3-\tfrac{49}{24}\big)\Tr(aabcbc)\nonumber\\
&+\big(\tfrac98\zeta_3-\tfrac{289}{576}\big)\Tr(aabccb)+\big(11\zeta_3-\tfrac{119}{18}\big)\Tr(abacbc)\nonumber\\
&-\big(\tfrac{11}{4}+\tfrac{15}{4}\zeta_3\big)\Tr(abcabc)\nonumber\\
&+\big(\tfrac{21}{8}\zeta_3-\tfrac{37}{32}\big)\Tr(aabc)\Tr(bc)+\big(\tfrac{23}{4}\zeta_3-\tfrac{187}{288}\big)\Tr(ab)\Tr(acbc)\nonumber\\
&+\big(\tfrac34\zeta_3-\tfrac{31}{576}\big)\Tr(ab)\Tr(ac)\Tr(bc)+\tfrac{127}{36}\lambda_{abcd}\Tr(abcd)\nonumber\\
&+\tfrac{253}{1728}\lambda_{acde}\lambda_{bcde}\Tr(ab)-\tfrac{7}{36}\lambda_{abcd}\lambda_{abef}\lambda_{cdef}\nonumber\\
&+\big(\tfrac76-\zeta_3\big)\Tr(abc)\Tr(abc)+\mathcal O(\epsilon^4).\label{eq:CT3}
\end{empheq}
The first two lines, equation~\eqref{eq:CT}, contain the free terms and the two- and three-loop invariants, and give $C_T$ through
$\mathcal O(\epsilon^2)$.  The remaining lines, equation~\eqref{eq:CT3}, are the four-loop invariants.  The four-loop basis has 22 independent
parity-even structures; the eleven not shown have vanishing coefficients.  They are the four disconnected $y^6$ traces such as $\Tr(aa)\Tr(bbcc)$, six
tadpole-type $\lambda$ contractions such as $\lambda_{abcc}\Tr(abdd)$, and $\lambda_{abcd}\Tr(ab)\Tr(cd)$.  The last term, \eqref{eq:oddT}, comes from fermion
loops with an odd number of Yukawa vertices.  It vanishes for every theory obtained from four dimensions, and for theories without a
four-dimensional counterpart it is fixed by $\mathcal N=1$ supersymmetry (sections~\ref{sec:dirac} and~\ref{sec:oddsusy}).

\paragraph{Conserved current.}
The four-loop part of $C_J$ is much longer than that of $C_T$, so here we give $C_J$ only through $\mathcal O(\epsilon^2)$.  At any fixed point,
\begin{equation}\label{eq:CJ}
\boxed{\;\begin{aligned}
\frac{C_J}{C_{J,\psi}}={}&\Tr(\Theta\Theta)+\frac{\theta_{ab}\theta_{ab}}{d-2}-\tfrac32\Tr(\Theta\Theta aa)\\
&-\Tr(\Theta\Theta aabb)-\tfrac{17}{2}\Tr(\Theta\Theta abab)-\tfrac58\Tr(\Theta\Theta abba)\\
&+\tfrac34\Tr(\Theta a \Theta abb)+\tfrac{15}{2}\Tr(\Theta a \Theta bab)+\tfrac32\Tr(\Theta ab \Theta ba)\\
&-\Tr(\Theta\Theta ab)\Tr(ab)+\tfrac32\Tr(\Theta a \Theta b)\Tr(ab)-\tfrac38\,\theta_{ae}\theta_{bf}\lambda_{abcd}\lambda_{efcd}+\mathcal O(\epsilon^3).
\end{aligned}\;}
\end{equation}
At $\mathcal O(\epsilon^3)$ the four-loop basis for $\langle JJ\rangle$ has 61 independent parity-even invariants (48 $y^6$, six $y^4\lambda$, four
$y^2\lambda^2$ and three $\lambda^3$ structures), of which 52 have non-zero coefficients.  They are given in equation~\eqref{eq:CJ3} of
appendix~\ref{app:CJ3}, and in machine-readable form in the repository~\cite{yukawactcj}.  As for $C_T$, its last two terms, \eqref{eq:oddJ}, contribute only to
theories without a four-dimensional counterpart.  Because of the Ward identity \eqref{eq:ward}, the $\langle JJ\rangle$
invariants are not all independent; the bases of \eqref{eq:CJ} and \eqref{eq:CJ3} are independent modulo the relations of appendix~\ref{app:ward}.

\paragraph{Comparison with known results.}
The $\Tr(aa)$ and $\lambda\cdot\lambda$ terms of \eqref{eq:CT} reproduce known results \cite{Hathrell1,JackOsborn,OS2016,Diab}, and the
$\Tr(\Theta\Theta aa)$ and $\theta\theta\lambda\lambda$ terms of \eqref{eq:CJ} reduce to the large-$N$ Gross--Neveu result of \cite{Diab} and the $O(n)$
result of \cite{Petkou}.  The $\lambda^3$ coefficient of \eqref{eq:CT3} agrees with the value $-\frac7{36}$ that \cite{OS} inferred indirectly;
given the $\lambda\cdot\lambda$ coefficient and the two-loop $\lambda_*$, it is fixed by the Ising $\mathcal O(\epsilon^3)$ result \cite{GKSS,HvL,IK}
(appendix~\ref{sec:checks}).  The three-loop Yukawa coefficients in \eqref{eq:CT} and \eqref{eq:CJ}, and all $\mathcal O(\epsilon^3)$ coefficients
that involve Yukawa couplings, are new.

\paragraph{Scheme dependence.}
The values of $C_T$ and $C_J$ at a fixed point are scheme independent, but the individual coefficients need not be.  A coupling redefinition
$y\to y+\mathcal O(y^3)$, $\lambda\to\lambda+\mathcal O(\lambda^2,\lambda y^2,y^4)$, which relates one scheme to another, changes $\Tr(y_*^ay_*^a)$ at
$\mathcal O(\epsilon^2)$ and $\lambda_*\cdot\lambda_*$ at $\mathcal O(\epsilon^3)$.  The contributions of the lower invariants then shift at these orders, and
the coefficients of the three-loop Yukawa invariants and of the four-loop invariants change to compensate.  Our values of these coefficients
therefore refer to fixed points of the $\overline{\rm MS}$ $\beta$-functions of \cite{JOS}.  The two-loop coefficients and those of
$\lambda\cdot\lambda$ and $\theta\theta\lambda\lambda$ are the same in every such scheme.

Equations \eqref{eq:CT} and \eqref{eq:CJ} hold only at fixed points.  Away from a fixed point the coefficients are ambiguous by multiples of
$\Tr(y^a\beta_y^a)$ and similar terms, where $\beta_y$ is the full $\beta$-function \eqref{eq:rg}, and we have chosen the representatives in which all
coefficients are $\epsilon$-independent.  The renormalized expressions before this reduction are given in the accompanying repository~\cite{yukawactcj}.

\paragraph{Sign of the leading corrections.}
Suppose the couplings are real, with real symmetric $y^a$ and real $\lambda_{abcd}$, as in the $4-\epsilon$ continuation of a unitary theory.  We
show that, at any fixed point, $C_T$ and $C_J$ then lie below their free values at the first order in $\epsilon$ at which they differ from them.
For $C_J$ this requires a mild non-degeneracy condition, stated below.

For $C_T$, both $\Tr(aa)$ and $\lambda\cdot\lambda$ in \eqref{eq:CT} are sums of squares.  If $Y_1\neq0$, the $\mathcal O(\epsilon)$ term
$-\frac54\,\epsilon\Tr(Y_1^aY_1^a)$ is negative.  If $Y_1=0$, the term $-\frac\epsilon2y^a$ in \eqref{eq:rg} forces $y_*=0$ at every order, and the
leading term is $-\frac5{36}\,\epsilon^2L_1\cdot L_1$, which is negative unless the fixed point is free.

For $C_J$, $\Tr(\Theta \Theta aa)=\Tr\big((y^a\Theta)^\dagger(y^a\Theta)\big)\ge0$, so the $\mathcal O(\epsilon)$ term
$-\frac32\,\epsilon\Tr(\Theta\Theta Y_1^aY_1^a)$ is non-positive.  If it vanishes, then $Y_1^a\Theta=0$, and hence also $\Theta Y_1^a=(Y_1^a\Theta)^\dagger=0$,
for every $a$.  Every Yukawa word in \eqref{eq:CJ} contains $\Theta$ next to a Yukawa matrix and therefore vanishes at $\mathcal O(\epsilon^2)$; this
includes the $\mathcal O(\epsilon^2)$ part $\Tr(\Theta\Theta Y_1^aY_2^a)+\Tr(\Theta\Theta Y_2^aY_1^a)$ of $\Tr(\Theta\Theta aa)$.  The
$\mathcal O(\epsilon^2)$ term is then $-\frac38\,\epsilon^2\theta_{ae}\theta_{bf}L_{1,abcd}L_{1,efcd}$.  By the last relation of appendix~\ref{app:ward},
$\theta_{ae}\theta_{bf}\lambda_{abcd}\lambda_{efcd}=\frac13\,\mathrm{tr}(M\Lambda)$ with $M_{ab}=(\theta \theta^{\rm T})_{ab}$ and
$\Lambda_{ab}=\lambda_{acdf}\lambda_{bcdf}$.  Both matrices are positive semi-definite, so this term is also non-positive.  If it vanishes as well, then
$M\Lambda=0$ at $\lambda=L_1$.  Together with $\theta_{ab}Y_1^b=0$, which follows from \eqref{eq:ward}, this means that every field charged under the
symmetry is decoupled at the one-loop fixed point.  Provided the one-loop stability matrix has no zero eigenvalue in the charged directions, the
charged sector then stays free at higher orders, and $C_J$ keeps its free value.  

\subsection{Two-coupling families}\label{sec:famres}
For the families \eqref{eq:family}, every invariant in \eqref{eq:CT} and \eqref{eq:CJ} reduces to $n_s$, $N$ and the constants \eqref{eq:tid}; for
example, $\Tr(y^ay^ay^by^b)=N n_s^2\alpha^2y^4$ and $\Tr(y^ay^by^ay^b)=N n_s\alpha\beta\,y^4$.  Reducing the one- and two-loop $\beta$-functions of
\cite{JOS} in the same way gives the fixed point
\begin{equation}
y_*^2=\frac{\epsilon}{2K}-\Big(\frac{c_5}{4K^2}+\frac{c_3\,\ell}{2K}+c_1\ell^2\Big)\frac{\epsilon^2}{K}+\mathcal O(\epsilon^3),\qquad \lambda_*=\ell\,\epsilon+\mathcal O(\epsilon^2),
\end{equation}
with $K$ as in \eqref{eq:fp1} and
\begin{align}
c_5&=2\gamma-2\beta^2-2n_s\alpha\beta-\tfrac32N\alpha\beta-\tfrac32Nn_s\alpha^2-\tfrac14n_s^2\alpha^2,\nonumber\\
c_3&=-2(2n_s\alpha+\beta),\qquad c_1=\tfrac14(n_s+2),\nonumber\\
\ell&=\frac{1}{2(n_s+8)}\Big[1-\frac{N\alpha}{K}+\sqrt{\Big(1-\frac{N\alpha}{K}\Big)^2+\frac{4(n_s+8)N\alpha(2n_s\alpha+\beta)}{(n_s+2)K^2}}\;\Big].
\end{align}
The one-loop $\beta_\lambda$ has two roots, which differ by the sign of the square root in $\ell$.  We give the one with the positive sign, the
stable branch $\lambda_+$, which we use throughout.  Substituting into \eqref{eq:CT} and \eqref{eq:CJ} gives
\begin{align}
\frac{C_T}{C_{T,\rm free}}&=1-\frac{\tfrac54 n_s\alpha N\,y_*^2+\tfrac{1}{48}n_s\alpha N(8N\alpha+11n_s\alpha+152\beta)\frac{\epsilon^2}{4K^2}+\tfrac{5}{12}n_s(n_s+2)\ell^2\epsilon^2}
{n_s+\tfrac12(d-1)N}+\mathcal O(\epsilon^3),\label{eq:famCT}\\
\frac{C_J}{C_{J,\rm free}}&=1-\tfrac32n_s\alpha\,y_*^2+\tfrac18 n_s\alpha\,(5n_s\alpha-8\beta+4N\alpha)\frac{\epsilon^2}{4K^2}+\mathcal O(\epsilon^3),\label{eq:famCJ}
\end{align}
In the $\mathcal O(\epsilon)$ terms we keep the full $y_*^2$ above; in the $\mathcal O(\epsilon^2)$ terms we have used $y_*^4=\epsilon^2/(4K^2)+\mathcal O(\epsilon^3)$.
The expression \eqref{eq:famCJ} for $C_J$ holds for any current acting on the fermions only ($\theta=0$), for instance fermion number.  With the
constants of table~\ref{tab:families}, equation \eqref{eq:famCT} gives $C_T$ for the chiral Ising, XY and Heisenberg models of the next subsection,
and equation \eqref{eq:famCJ} gives the chiral Ising $C_J$ \eqref{eq:isingCJ} and the fermion-number $C_J$ of table~\ref{tab:sigma}.

\subsection{Chiral Ising, XY and Heisenberg models}\label{sec:models}
We now evaluate these results for the chiral Ising, XY and Heisenberg models, on the stable branch $\lambda_+$.  For chiral Ising,
\eqref{eq:famCT} and \eqref{eq:famCJ} give
\begin{align}
\text{chiral Ising:}\quad &\frac{C_T}{C_{T,\rm free}}=1-\frac{5N\epsilon}{2(N+6)(3N+2)}
-\frac{P(N)+S\,R(N)}{648(N+6)^3(3N+2)^2}\,\epsilon^2,\label{eq:isingCT}\\
&\frac{C_J}{C_{J,\rm free}}=1-\frac{3\epsilon}{2(N+6)}+\frac{37N^2-327N-1044-S\,(N+66)}{72(N+6)^3}\,\epsilon^2,\label{eq:isingCJ}\\
&P=2253N^4+62153N^3+217302N^2+93672N+4320,\nonumber\\
&R=15N^3+2980N^2+3060N+720,\qquad S=\sqrt{N^2+132N+36}.\nonumber
\end{align}
For chiral XY and Heisenberg, $C_T$ again follows from \eqref{eq:famCT} with the constants of table~\ref{tab:families}.  Their currents also act on
the scalars, so $C_J$ is not covered by \eqref{eq:famCJ} and we evaluate \eqref{eq:CJ} directly.  For XY we take the $U(1)$ under which the fermions
have charge 1 and the scalars charge 2, and find
\begin{equation}
\frac{C_J}{C_{J,\rm free}}=1-\frac{3N\epsilon}{(N+4)^2}-\frac{U_2-V_2\,S_2}{100(N+4)^4}\,\epsilon^2,
\end{equation}
For Heisenberg we take the $SU(2)$ generator $\sigma_3/2$, and find
\begin{equation}\label{eq:heisCJ}
\frac{C_J}{C_{J,\rm free}}=1-\frac{9N\epsilon}{2(N+2)(N+4)}-\frac{3\,(U_3-V_3\,S_3)}{968(N+2)^3(N+4)^2}\,\epsilon^2,
\end{equation}
with
\[
\begin{gathered}
U_2=503N^3+3550N^2-2464N+384,\qquad V_2=3(N^2-76N-32),\\
U_3=1457N^4+7993N^3-30816N^2-169360N+640,\qquad V_3=5(N+4)(N^2-258N-16),\\
S_2=\sqrt{N^2+152N+16},\qquad S_3=\sqrt{N^2+172N+4}.
\end{gathered}
\]
Table~\ref{tab:models2} gives the numerical coefficients.\footnote{For chiral Ising at $N=2$ we take the theory with a $U(1)$
symmetry, a two-component Dirac fermion coupled to a real scalar, so that $C_J$ is defined.  It has no four-dimensional counterpart:
$\Tr(y^ay^by^c)\neq0$, and its $C_T$ includes the odd fermion loops \eqref{eq:oddT}, which add $\big(\frac7{3072}-\frac{\zeta_3}{512}\big)\epsilon^3$ and
change the $\epsilon^3$ coefficient from $+0.01156$ to $+0.01149$.  Its $C_J$ is unchanged, since $\Tr(\Theta y^ay^by^c)=0$.  The theory obtained from a
four-dimensional Majorana fermion, two two-component Majorana fermions with opposite signs of $y$, has the same even invariants and hence the same
coefficients without this term, but no continuous flavor symmetry.  At $N=4$ and $8$ the entries are for the theories obtained from four dimensions
(graphene for $N=8$), for which the odd invariants vanish.}  The
$\epsilon^3$ terms use \eqref{eq:CT3} and \eqref{eq:CJ3} together with the fixed point to $\mathcal O(\epsilon^3)$, obtained from the three-loop
$\beta_y$ (appendix~\ref{app:fp3}).
\begin{table}[htbp]
\centering\small\setlength{\tabcolsep}{5pt}
\begin{tabular}{llrrrrrr}\toprule
 & & \multicolumn{3}{c}{$C_T/C_{T,\rm free}-1$} & \multicolumn{3}{c}{$C_J/C_{J,\rm free}-1$}\\
model & $N$ & $\epsilon$ & $\epsilon^2$ & $\epsilon^3$ & $\epsilon$ & $\epsilon^2$ & $\epsilon^3$\\ \midrule
chiral Ising & 1 & $-0.0714$ & $-0.0842$ & $+0.0109$ & --- & --- & ---\\
 & 2 & $-0.0781$ & $-0.0906$ & $+0.0115$ & $-0.1875$ & $-0.0742$ & $+0.0504$\\
 & 4 & $-0.0714$ & $-0.0779$ & $+0.0113$ & $-0.1500$ & $-0.0479$ & $+0.0404$\\
 & 8 & $-0.0549$ & $-0.0527$ & $+0.0127$ & $-0.1071$ & $-0.0193$ & $+0.0314$\\
\midrule
chiral XY & 1 & $-0.1429$ & $-0.0792$ & $-0.0435$ & $-0.1200$ & $-0.0983$ & $-0.0691$\\
 & 2 & $-0.1667$ & $-0.1250$ & $+0.0157$ & $-0.1667$ & $-0.1806$ & $-0.0028$\\
 & 4 & $-0.1563$ & $-0.1331$ & $+0.0354$ & $-0.1875$ & $-0.2534$ & $+0.0195$\\
 & 8 & $-0.1190$ & $-0.0975$ & $+0.0368$ & $-0.1667$ & $-0.2544$ & $+0.0215$\\
\midrule
chiral Heisenberg & 1 & $-0.2778$ & $+0.2466$ & $-1.5108$ & $-0.3000$ & $+0.4559$ & $-2.3661$\\
 & 2 & $-0.3125$ & $-0.0383$ & $-0.1517$ & $-0.3750$ & $+0.1034$ & $-0.4168$\\
 & 4 & $-0.2778$ & $-0.1711$ & $+0.0855$ & $-0.3750$ & $-0.1823$ & $+0.0773$\\
 & 8 & $-0.2000$ & $-0.1404$ & $+0.0806$ & $-0.3000$ & $-0.2439$ & $+0.0955$\\
\bottomrule\end{tabular}
\caption{Coefficients of $\epsilon$, $\epsilon^2$, $\epsilon^3$ in $C_T/C_{T,\rm free}-1$ and $C_J/C_{J,\rm free}-1$ for the chiral models on the stable branch $\lambda_+$.
In the Ising model, the current is any fermion-only flavor current. In the XY model, the current is the $U(1)$ with fermion charge 1 and scalar charge 2. In the Heisenberg model, the current is the $SU(2)$ generator $\sigma_3/2$.
Chiral Ising at $N=1$ has no continuous flavor symmetry, so there is no $C_J$ to compute.  Its $\epsilon^3$ entry includes the odd fermion loops
\eqref{eq:oddT}, which change it by $-4\times10^{-5}$.  The $N=1$ entries of the chiral XY and Heisenberg models are formal continuations in $N$ and do not
include them.}
\label{tab:models2}
\end{table}

Two of the entries are supersymmetric fixed points, chiral Ising at $N=1$ and chiral XY at $N=2$, which we discuss in section~\ref{sec:susy}.
At the latter, the $C_J$ entries differ from the $\mathcal N=2$ value \eqref{eq:susy} beyond $\mathcal O(\epsilon)$ because $C_{J,\rm free}$ here uses the
charges $(1,2)$ rather than the free $R$-charges; section~\ref{sec:Rcurrent} derives the exact factor relating them.

\subsection{Gross--Neveu--Yukawa model at large \texorpdfstring{$N$}{N}}\label{sec:largeN}
For the Gross--Neveu--Yukawa model (the chiral Ising model) we write $C_T=C_{T0}\,(1+C_{T1}/N+C_{T2}/N^2+\dots)$ and
$C_J=C_{J0}\,(1+C_{J1}/N+C_{J2}/N^2+\dots)$, where $C_{T0}$ and $C_{J0}$ are the free-fermion values, as in \cite{Diab}.  The $1/N^2$ coefficients are
\begin{align}
C_{T2}&=5\,\epsilon-\frac{307}{72}\,\epsilon^2-\Big(\frac{89}{48}+\frac{41\,\zeta_3}{12}\Big)\epsilon^3+\mathcal O(\epsilon^4),\nonumber\\
C_{J2}&=9\,\epsilon-\frac{123}{8}\,\epsilon^2+\Big(\frac{9\,\zeta_3}{4}-\frac{57}{16}\Big)\epsilon^3+\mathcal O(\epsilon^4).
\end{align}
The $\epsilon$ and $\epsilon^2$ terms follow by expanding \eqref{eq:isingCT} and \eqref{eq:isingCJ} in $1/N$.  For the $\epsilon^3$ terms we solve the
three-loop fixed point as a series in $1/N$ and expand \eqref{eq:CT3} and \eqref{eq:CJ3}, in which every invariant carries an exact power of $N$.  As
a check, the same expansion reproduces the known $C_{T1}$ and $C_{J1}$ of \cite{Diab} through $\epsilon^3$ (appendix~\ref{sec:checks}).
To our knowledge the $1/N^2$ coefficients of $C_T$ and $C_J$ have not been computed for the Gross--Neveu model; for the $O(N)$ model, $C_J$ is known
at this order \cite{AldayHenrikssonvanLoon}.

\section{Emergent supersymmetry in \texorpdfstring{$d=3$}{d=3}}\label{sec:susy}
Several of the fixed points above flow, in $d=3$, to theories with emergent supersymmetry \cite{Lee2007,PonteLee,GroverShengVishwanath,FGKT,LiendoRong,PrakashSinha}.
What can $C_T$ and $C_J$ tell us about them?  First, with four supercharges the $R$-current and the stress tensor lie in one multiplet, so there is an exact
relation between $C_T$ and $C_J^R$.  Checking this relation order by order tests the emergence of supersymmetry.  Second, our $d=3$ estimates can be compared
with exact results from localization and with the bootstrap.

We first give the supersymmetric points of the chiral models (section~\ref{sec:susychiral}) and identify the $R$-current at the $\mathcal N=2$ point
(section~\ref{sec:Rcurrent}).  We then turn to the $\mathcal N=1$ points of Liendo and Rong (section~\ref{sec:LR}), show that $\mathcal N=1$ supersymmetry fixes the contribution of
fermion loops with an odd number of Yukawa vertices (section~\ref{sec:oddsusy}), consider the matrix models of \cite{PrakashSinha}
(section~\ref{sec:matrix}).  Finally, we estimate $C_T$ and $C_J$ in $d=3$ for all these theories using Pad\'e resummation (section~\ref{sec:susyd3}).

\subsection{The chiral models}\label{sec:susychiral}
Chiral Ising at $N=1$ flows to a fixed point with emergent $\mathcal N=1$ supersymmetry,
and chiral XY at $N=2$ to the $\mathcal N=2$ Wess--Zumino (super-Ising) fixed point \cite{FGKT}.  There
\begin{align}
\frac{C_T}{C_{T,\rm free}}\Big|_{\mathcal N=2}&=1-\frac{\epsilon}{6}-\frac{\epsilon^2}{8}+\Big(\frac{\zeta_3}{12}-\frac{73}{864}\Big)\epsilon^3+\mathcal O(\epsilon^4)=\frac{C_J^R}{C^R_{J,\rm free}}\Big|_{\mathcal N=2},\nonumber\\
\frac{C_T}{C_{T,\rm free}}\Big|_{\mathcal N=1}&=1-\frac{\epsilon}{14}-\frac{33\,\epsilon^2}{392}+\Big(\frac{597\,\zeta_3}{9604}-\frac{14713}{230496}\Big)\epsilon^3+\mathcal O(\epsilon^4).\label{eq:susy}
\end{align}
The $\mathcal N=1$ series includes the odd fermion loops \eqref{eq:oddT}, which contribute $\big(\frac1{735}-\frac{2\zeta_3}{1715}\big)\epsilon^3$.
Here $C_J^R$ is the coefficient of the superconformal $R$-current of the interacting fixed point, and $C^R_{J,\rm free}$ is its value with the $R$-charges of
the free theory; section~\ref{sec:Rcurrent} identifies this current.
The equality of the two $\mathcal N=2$ ratios is required by superconformal symmetry.  It is not built into our calculation, since $C_T$ and $C_J^R$ come from
different diagrams, projectors and counterterms.  We have also built eight $\mathcal N=2$ Wess--Zumino models from their symmetry alone, without imposing
supersymmetry, and the identity holds at each of their fixed points through $\epsilon^3$ (appendix~\ref{app:chkCJ3}), so it is not specific to the chiral XY model.

\subsection{The \texorpdfstring{$U(1)$}{U(1)} current and the \texorpdfstring{$R$}{R}-current at the \texorpdfstring{$\mathcal N=2$}{N=2} point}\label{sec:Rcurrent}
The chiral XY model at $N=2$ has a single $U(1)$ symmetry, which acts with charge 1 on the fermion and charge 2 on the scalar.  At the
$\mathcal N=2$ fixed point it is the only continuous symmetry, so it must be proportional to the superconformal $R$-symmetry, and supersymmetry then
relates its $C_J$ to $C_T$.  One might therefore expect $C_J/C_{J,\rm free}=C_T/C_{T,\rm free}$ for this current.  Table~\ref{tab:models2} shows
that this fails at $\mathcal O(\epsilon^2)$.  The reason is that the free theory has a second $U(1)$ symmetry: at the free point the same current is
no longer proportional to the superconformal $R$-current, so $C_{J,\rm free}$ is not related to $C_{T,\rm free}$ in the same way.  In this section we
derive the exact factor between the two ratios, \eqref{eq:TIfactor}.  In $d=3$ this current is the electric current at the semimetal--superconductor
transition on the surface of a topological insulator, so the factor determines its conductivity (section~\ref{sec:applications}).

An $R$-symmetry is a symmetry that rotates the supercharges, $[R,Q_\alpha]=-Q_\alpha$.  It is not unique. If $R_0$ is an $R$-symmetry and $\mathcal F$
is a flavor symmetry, which commutes with the supercharges, then $R_0+t\mathcal F$ is also an $R$-symmetry for every $t$.  The superconformal algebra
$\mathfrak{osp}(2|4)$, however, contains exactly one of them, the superconformal $R$-symmetry.  Its current $J^R_\mu$ is the bottom component of the
stress-tensor multiplet, which also contains the supersymmetry currents and $T_{\mu\nu}$ \cite{DumitrescuSeiberg,ClossetEtAl}.  The superconformal
Ward identities fix the two-point function of this multiplet up to a single constant \cite{BKS2015}, so
\begin{equation}\label{eq:kappa}
C_J^R=\kappa\,C_T ,
\end{equation}
with $\kappa$ universal.  Flavor currents, by contrast, lie in linear multiplets, whose normalization is independent of $C_T$.  The relation
\eqref{eq:kappa} therefore holds for a $U(1)$ current if and only if the current is proportional to the superconformal $R$-current.  The same
structure holds for four supercharges in $d=4$, where $C_J^R\propto c$, and we assume it continues to hold for the entire range $2\le d\le4$, as in the check of
section~\ref{sec:susychiral}.

In superspace the fixed point is the Wess--Zumino model of one chiral superfield $\Phi=(\phi,\chi,F)$ with superpotential $W=\frac g3\Phi^3$.  Here
$\chi$ is the charge conjugate of the fermion $\psi$, so that $\phi$ and $\chi$ carry $U(1)$ charges $2$ and $-1$, and after eliminating $F$ the
interactions are $-|g|^2|\phi|^4-\big(g\,\phi\,\chi\chi+{\rm c.c.}\big)$.  Since the Grassmann coordinate  pairs with $Q_\alpha$, it carries
$R$-charge $+1$, and the components of a chiral superfield with $R$-charge $R_\Phi$ have
\begin{equation}\label{eq:Rcomp}
R[\phi]=R_\Phi,\qquad R[\chi]=R_\Phi-1,\qquad R[F]=R_\Phi-2
\end{equation}
(see, e.g., \cite{DumitrescuSeiberg,ClossetEtAl}).  The superpotential term is neutral only if $R[W]=2$, so $W\propto\Phi^3$ requires $R_\Phi=\frac23$.

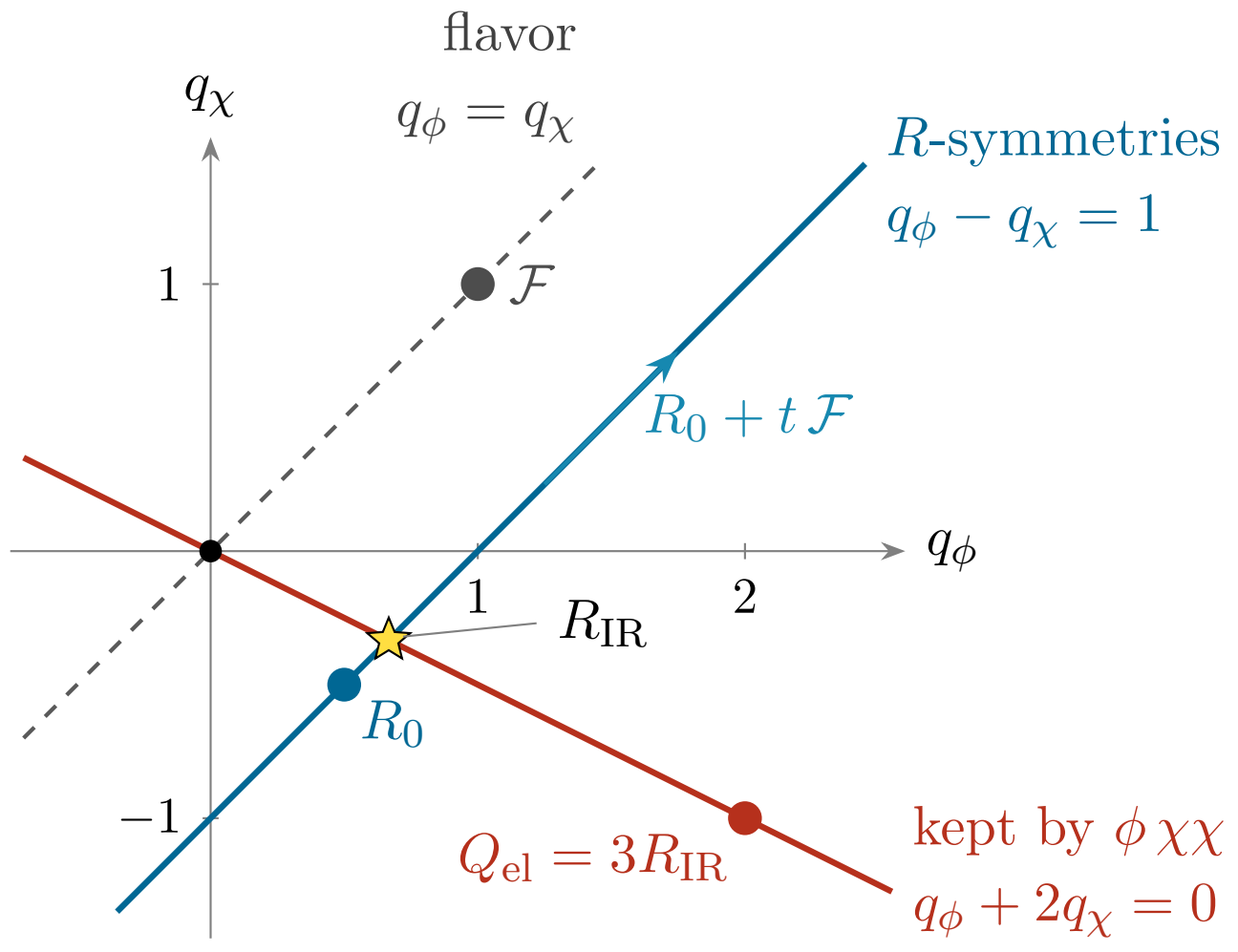
\begin{figure}[t]
\centering
\begin{tikzpicture}[>=Stealth,scale=1.85,font=\small]
\draw[->,gray] (-0.75,0)--(2.6,0) node[right,black]{$q_\phi$};
\draw[->,gray] (0,-1.45)--(0,1.55) node[above,black]{$q_\chi$};
\foreach \x in {1,2}{\draw[gray] (\x,0.03)--(\x,-0.03) node[below,black,font=\footnotesize]{$\x$};}
\foreach \y in {-1,1}{\draw[gray] (0.03,\y)--(-0.03,\y) node[left,black,font=\footnotesize]{$\y$};}
\draw[thick,dashed,gray!70!black] (-0.7,-0.7)--(1.45,1.45);
\node[gray!50!black,above left,align=right] at (1.45,1.45) {flavor\\ $q_\phi=q_\chi$};
\draw[very thick,MidnightBlue] (-0.35,-1.35)--(2.45,1.45);
\node[MidnightBlue,right,align=left] at (2.45,1.38) {$R$-symmetries\\ $q_\phi-q_\chi=1$};
\draw[very thick,BrickRed] (-0.7,0.35)--(2.55,-1.275);
\node[BrickRed,right,align=left] at (2.55,-1.2) {kept by $\phi\,\chi\chi$\\ $q_\phi+2q_\chi=0$};
\draw[->,thick,MidnightBlue!70] (1.25,0.25)--(1.75,0.75) node[midway,right=2pt]{$R_0+t\,\mathcal F$};
\fill (0,0) circle (1.2pt);
\fill[MidnightBlue] (0.5,-0.5) circle (1.8pt) node[below right=-1pt]{$R_0$};
\fill[gray!60!black] (1,1) circle (1.8pt) node[right=2pt]{$\mathcal F$};
\node[star,star points=5,star point ratio=2.25,fill=Goldenrod,draw,inner sep=1.4pt] at (0.6667,-0.3333) {};
\draw[gray] (0.72,-0.32)--(1.22,-0.27) node[right,black]{$R_{\rm IR}$};
\fill[BrickRed] (2,-1) circle (1.8pt) node[below left=-1pt]{$Q_{\rm el}=3R_{\rm IR}$};
\end{tikzpicture}
\caption{The plane of internal $U(1)$ symmetries of the free chiral multiplet, labelled by the charges $(q_\phi,q_\chi)$.  Flavor symmetries lie on the dashed
line and $R$-symmetries on the blue line, which the family $R_0+t\,\mathcal F$ sweeps out; $R_0$ is the point for $d=3$.  The cubic superpotential keeps only
the red line.  Its intersection with the blue line is $R_{\rm IR}$, the unique $R$-symmetry of the interacting theory.  The electric charge also lies on the
red line, at three times $R_{\rm IR}$.}
\label{fig:chargeplane}
\end{figure}

At the free point ($g=0$)  $\phi$ and $\chi$ decouple, and each kinetic term is separately invariant under an independent phase rotation of its own field.
The free theory therefore has a two-dimensional space of $U(1)$ symmetries, labelled by the charges $(q_\phi,q_\chi)$ they assign, as shown in
figure~\ref{fig:chargeplane}.
Flavor symmetries, by definition, commute with the supercharges, so they must give $\phi$ and $\chi$ equal charges; they lie on the line $q_\phi=q_\chi$, spanned by $\mathcal F=(1,1)$.  By \eqref{eq:Rcomp}, an $R$-symmetry gives $\chi$ one unit less charge than
$\phi$, so the $R$-symmetries lie on the parallel line $q_\phi-q_\chi=1$, and each of them is $R_0+t\mathcal F$ for some $t$, where $R_0$ is the
superconformal $R$-symmetry of the free theory.  To determine $R_0$, we use the fact that  $\Delta=\frac{d-1}{2}\,r$ for chiral primaries. Since $\Delta_\phi=\frac{d-2}{2}$,
$R_0$ assigns $r_\phi=\frac{d-2}{d-1}$ and $r_\chi=r_\phi-1=-\frac1{d-1}$, which is $(\frac12,-\frac12)$ in $d=3$.

The interaction $g\,\phi\,\chi\chi$ is invariant only under the $U(1)$ symmetries on the line $q_\phi+2q_\chi=0$, and it breaks the flavor symmetry $\mathcal F$, under
which $W$ has charge 3.  The only $R$-symmetry it preserves is the intersection of this line with the line of $R$-symmetries. We call this $R$-symmetry
$R_{\rm IR}$; it has $r_\phi=\frac23$ in any $d$, so
\begin{equation}
R_{\rm IR}=R_0+t\,\mathcal F,\qquad t=\frac23-\frac{d-2}{d-1}=\frac{4-d}{3(d-1)} ,
\end{equation}
and $t=\frac16$ in $d=3$.  $R_{\rm IR}$ is the only continuous internal symmetry of the interacting theory, so it is the superconformal $R$-symmetry of the interacting fixed point.

To summarize, the charges of $\phi$ and $\chi$ under these three symmetries are
\begin{center}\small
\begin{tabular}{lcccc}\toprule
 & $Q_{\rm el}$ & $R_0$ & $\mathcal F$ & $R_{\rm IR}$\\\midrule
$\phi$ & $2$ & $\frac{d-2}{d-1}$ & $1$ & $\frac23$\\[2pt]
$\chi$ & $-1$ & $-\frac{1}{d-1}$ & $1$ & $-\frac13$\\\bottomrule
\end{tabular}
\end{center}
where $Q_{\rm el}$ is the $U(1)$ charge of the chiral XY model.  An identity between charges, which holds along the whole flow, is
\begin{equation}\label{eq:Qel}
Q_{\rm el}=3R_{\rm IR}=3R_0+\frac{4-d}{d-1}\,\mathcal F .
\end{equation}

Which $R$-symmetry is superconformal changes along the flow.  At the interacting fixed point it is $R_{\rm IR}$, so the current of $Q_{\rm el}$
is three times the superconformal $R$-current, and \eqref{eq:kappa} gives $C_J=9\kappa\,C_T$.  At the free point the superconformal
$R$-symmetry is $R_0$, and by \eqref{eq:Qel} the same current is $3J^{R_0}+\frac{4-d}{d-1}\,J^{\mathcal F}$, where $J^{\mathcal F}$, the current of
$\mathcal F$, is conserved only at the free point.  The two terms do not mix: the two-point function of $J^{R_0}$ with $J^{\mathcal F}$ is
proportional to $r_\phi/(d-2)+r_\chi=0$, because, per unit charge, $C_J$ of a free complex scalar is $1/(d-2)$ times $C_J$ of a free two-component
fermion.  Hence
\begin{equation}
C_{J,\rm free}=9\,C^{R_0}_{J,\rm free}+\Big(\frac{4-d}{d-1}\Big)^2C^{\mathcal F}_{J,\rm free}.
\end{equation}
Since $R_0$ is the superconformal $R$-symmetry of the free theory, \eqref{eq:kappa} applies to its current there, $C^{R_0}_{J,\rm free}=\kappa\,C_{T,\rm free}$,
whereas the $J^{\mathcal F}$ term is not related to $C_{T,\rm free}$.  Writing $C_J^\chi$ for $C_J$ of the fermion per unit charge, the charges in the
table give $C^{R_0}_{J,\rm free}=C_J^\chi/(d-1)$, $C^{\mathcal F}_{J,\rm free}=C_J^\chi\,(d-1)/(d-2)$ and $C_{J,\rm free}=C_J^\chi\,(d+2)/(d-2)$.
Dividing $C_J=9\kappa\,C_T$ by $C_{J,\rm free}$ and using $C^{R_0}_{J,\rm free}=\kappa\,C_{T,\rm free}$ gives
\begin{equation}\label{eq:TIfactor}
\frac{C_J}{C_{J,\rm free}}=\frac{9\,C^{R_0}_{J,\rm free}}{C_{J,\rm free}}\,\frac{C_T}{C_{T,\rm free}}
=\frac{9(d-2)}{(d-1)(d+2)}\,\frac{C_T}{C_{T,\rm free}} .
\end{equation}
The prefactor is $1-\epsilon^2/18-\epsilon^3/36+\mathcal O(\epsilon^4)$, and $\frac9{10}$ in $d=3$, where the $J^{\mathcal F}$ term is a tenth of
$C_{J,\rm free}$.  The $J^{\mathcal F}$ term has coefficient $\big(\frac{4-d}{d-1}\big)^2=9t^2$, and $t$ vanishes in $d=4$, where the free chiral
multiplet already has $r_\phi=\frac23$.  The prefactor therefore differs from 1 first at $\mathcal O(\epsilon^2)$.  Our $\epsilon$-expansion confirms \eqref{eq:TIfactor} independently: we computed $C_J$ directly from the $U(1)$
current, without using supersymmetry, and the chiral XY entries of table~\ref{tab:models2} at $N=2$ equal this prefactor times \eqref{eq:susy} through
$\mathcal O(\epsilon^3)$.

\subsection{The \texorpdfstring{$\mathcal N=1$}{N=1} fixed points of Liendo and Rong}\label{sec:LR}
Liendo and Rong \cite{LiendoRong} classified the $\mathcal N=1$ fixed points with $N_\Phi$ real superfields.  As for the $\mathcal N=2$ models of
section~\ref{sec:susychiral}, we build each theory from its symmetry alone, without imposing supersymmetry, and recover the supersymmetric fixed point, with the
leading anomalous dimension $\gamma_\Phi$ of the superfield as in \cite{LiendoRong}.  The $S_3$ super-Potts point coincides with the $\mathcal N=2$ point above.
Table~\ref{tab:LR} gives $C_T$ at the other points, including the odd fermion loops \eqref{eq:oddT}.  The $O(3)/\mathbb Z_2$ point is also the supersymmetric point of the $SO(3)$ matrix model of
section~\ref{sec:matrix}, where we give its $C_J$.
\begin{table}[htbp]
\centering\small
\begin{tabular}{lll}\toprule
fixed point & $\gamma_\Phi/\epsilon$ & $C_T/C_{T,\rm free}$\\\midrule
$S_4$ super-Potts ($N_\Phi=3$) & $1/10$ & $1-0.1\,\epsilon-0.101\,\epsilon^2+0.027074\,\epsilon^3$\\
$S_5$ super-Potts ($N_\Phi=4$) & $3/34$ & $1-0.088235\,\epsilon-0.094876\,\epsilon^2+0.014636\,\epsilon^3$\\
$O(3)/\mathbb Z_2$ ($N_\Phi=5$) & $7/30$ & $1-0.233333\,\epsilon-0.157889\,\epsilon^2-0.035194\,\epsilon^3$\\\bottomrule
\end{tabular}
\caption{$C_T$ at the $\mathcal N=1$ fixed points of Liendo and Rong \cite{LiendoRong} with $N_\Phi$ real superfields.  The $S_3$ super-Potts point is the
$\mathcal N=2$ point of \eqref{eq:susy}.}
\label{tab:LR}
\end{table}

\subsection{\texorpdfstring{$\mathcal N=1$}{N=1} supersymmetry and the odd fermion loops}\label{sec:oddsusy}
Three-dimensional $\mathcal N=1$ supersymmetry has no $R$-symmetry, but the stress tensor still lies in a multiplet with the supersymmetry current
$S_{\mu\alpha}$, so that the coefficient $C_S$ of its two-point function satisfies
\begin{equation}\label{eq:CSCT}
\frac{C_S}{C_{S,\rm free}}=\frac{C_T}{C_{T,\rm free}}\qquad(d=3).
\end{equation}
Likewise, a flavor current $J_\mu$, generated by $\theta_{ab}$ acting alike on $\phi_a$ and $\psi_a$, lies in a multiplet whose lowest component is the spinor
$j_\alpha=\theta_{ab}\,\phi_a\psi_{b\alpha}$, of dimension $\frac32$, so that $C_j/C_{j,\rm free}=C_J/C_{J,\rm free}$.  We compute both in the Wess--Zumino theories,
with superpotential $W=\frac16h_{abc}\Phi_a\Phi_b\Phi_c$, so that $y^a_{ij}=h_{aij}$ and $\lambda_{abcd}=h_{abe}h_{cde}+h_{ace}h_{bde}+h_{ade}h_{bce}$.
The operators $S_\mu$, $\bar S_\nu$, $j$ and $\bar j$ in the conventions of \eqref{eq:action}, and the projections we use, are given in
appendix~\ref{app:supercurrent}.

Both $S$ and $j$ contain a single fermion field.  In any diagram contributing to the two-point function, the fermion line through them  must be open, and any closed fermion loops contain only Yukawa
vertices.  A closed loop with an odd number of Yukawa vertices gives one $\tilde\epsilon$-tensor, which survives integration only if at least four scalar
lines leave the loop, so it needs at least five Yukawa vertices; a parity-even result needs a second $\tilde\epsilon$-tensor, from the open line or another
loop.  This first happens at five loops, at $\mathcal O(\epsilon^4)$.  Up to $\mathcal O(\epsilon^3)$, $C_S$ and $C_j$ are
therefore unambiguous, and may be obtained using the usual four-dimensional Dirac algebra.

Table~\ref{tab:oddsusy} compares $C_S$ with $C_T$ and $C_j$ with $C_J$.  Through $\mathcal O(\epsilon^2)$, where the four-dimensional Dirac
algebra is exact, both relations hold at every point.  At $\mathcal O(\epsilon^3)$ they hold at every point once the odd fermion loops are included
with the continuation of section~\ref{sec:dirac}, and fail without them or with any of the other continuations (appendix~\ref{app:odd},
table~\ref{tab:oddcont}).  At leading order in the couplings, this continuation gives
\begin{align}
\Delta\frac{C_T}{C_{T,S}}&=\Big(\frac76-\zeta_3\Big)\Tr(y^ay^by^c)\Tr(y^ay^by^c),\label{eq:oddT}\\
\Delta\frac{C_J}{C_{J,\psi}}&=-3\,\Tr(\Theta y^ay^by^c)\Tr(\Theta y^ay^by^c)+(2-6\zeta_3)\,\Tr(\Theta y^ay^by^c)\Tr(\Theta y^by^cy^a),\label{eq:oddJ}
\end{align}
the last terms of \eqref{eq:CT3} and \eqref{eq:CJ3}.  Since $\langle S\bar S\rangle$ and $\langle j\bar j\rangle$ share no diagrams, projectors or
Fourier transforms with $\langle TT\rangle$ and $\langle JJ\rangle$, these relations also test our four-loop coefficients
(appendix~\ref{sec:checks}).

\begin{table}[htbp]
\centering\small\setlength{\tabcolsep}{6pt}
\begin{tabular}{lrrr}\toprule
fixed point & $C_S/C_{S,\rm free}$ & $C_T/C_{T,\rm free}$ & $C_T/C_{T,\rm free}$, no odd loops\\\midrule
$\mathcal N=1$ super-Ising & $+0.0108899$ & $+0.0108899$ & $+0.0109311$\\
$S_3$ super-Potts ($\mathcal N=2$) & $+0.0156807$ & $+0.0156807$ & $+0.0156807$\\
$S_4$ super-Potts & $+0.0270745$ & $+0.0270745$ & $+0.0271028$\\
$S_5$ super-Potts & $+0.0146359$ & $+0.0146359$ & $+0.0146704$\\
$O(3)/\mathbb Z_2$ & $-0.0351945$ & $-0.0351945$ & $-0.0351284$\\\midrule
 & $C_j/C_{j,\rm free}$ & $C_J/C_{J,\rm free}$ & $C_J/C_{J,\rm free}$, no odd loops\\\midrule
$O(3)/\mathbb Z_2$ & $+0.0252812$ & $+0.0252812$ & $-0.0943582$\\
$SU(3)$ adjoint [susy] & $-0.0605649$ & $-0.0605649$ & $-0.1886491$\\\bottomrule
\end{tabular}
\caption{Coefficients of $\epsilon^3$ at the $\mathcal N=1$ fixed points.  $C_S$ and $C_j$ have no odd fermion loops at this order.  $C_T$ and $C_J$
include them with the continuation of section~\ref{sec:dirac}, \eqref{eq:oddT} and \eqref{eq:oddJ}; the last column omits them.  At the $S_3$ point
$\Tr(y^ay^by^c)=0$, so there are no odd loops.  The $C_J$ entries are for a generator of $O(3)$ and of $SU(3)$.}
\label{tab:oddsusy}
\end{table}

\subsection{Matrix Gross--Neveu--Yukawa models}\label{sec:matrix}
Consider next the Gross--Neveu--Yukawa models with matrix degrees of freedom of \cite{PrakashSinha}.  In this subsection $N$ denotes the rank of $SU(N)$ or
$SO(N)$, not $\Tr\mathbf 1$.  In the first model the real pseudo-scalars and the Majorana fermions both transform in the adjoint of $SU(N)$; in the second both
are real symmetric traceless matrices ($S^2$) of $SO(N)$; in the third the scalars are in $S^2$ and the fermions in the adjoint ($A^2$) of $SO(N)$.  Each
model has a single Yukawa coupling $y$ and the quartic couplings $g_1\,\mathrm{tr}\,\phi^4+g_2\,(\mathrm{tr}\,\phi^2)^2$.  The large-$N$ limits of the three
models are perturbatively equivalent and dominated by planar diagrams.  The first two models have a supersymmetric fixed point for all $N$, which is infrared
stable for $N$ large enough, and the third has a non-supersymmetric fixed point that becomes equivalent to it at large $N$ \cite{PrakashSinha}.

The three-loop $\beta$-functions of \cite{PrakashSinha} were obtained from those of \cite{JOS}, and appendix~A of \cite{PrakashSinha} gives the tensors
$y^a_{ij}$ and $\lambda_{abcd}$ of each model in the conventions of \cite{JOS}, with the fermion number continued to $N_f=\frac12$.  We build these tensors explicitly, insert the fixed-point couplings of \cite{PrakashSinha}, which are known in closed form through
$\mathcal O(\epsilon^3)$ for the supersymmetric points and for the fixed points at $N=3$, and evaluate \eqref{eq:CT}, \eqref{eq:CJ}, \eqref{eq:CT3} and \eqref{eq:CJ3}.
For $C_J$ we take a generator of $SU(N)$ or $SO(N)$, which acts on both the scalars and the fermions.  Table~\ref{tab:matrix} gives the results.
\begin{table}[htbp]
\centering\small\setlength{\tabcolsep}{4pt}
\begin{tabular}{lrrrrrr}\toprule
 & \multicolumn{3}{c}{$C_T/C_{T,\rm free}-1$} & \multicolumn{3}{c}{$C_J/C_{J,\rm free}-1$}\\
fixed point & $\epsilon$ & $\epsilon^2$ & $\epsilon^3$ & $\epsilon$ & $\epsilon^2$ & $\epsilon^3$\\ \midrule
$SU(3)$ adjoint, [susy] & $-0.27778$ & $-0.20319$ & $-0.08956$ & $-0.55556$ & $-0.26749$ & $-0.06056$ \\
$SU(4)$ adjoint, [susy] & $-0.13636$ & $-0.11523$ & $+0.02000$ & $-0.27273$ & $-0.16228$ & $+0.03499$ \\
$SU(6)$ adjoint, [susy] & $-0.11111$ & $-0.10597$ & $-0.00809$ & $-0.22222$ & $-0.15638$ & $-0.02257$ \\
$SO(3)$ $S^2$, [susy] & $-0.23333$ & $-0.15789$ & $-0.03519$ & $-0.46667$ & $-0.19911$ & $+0.02528$ \\
$SO(4)$ $S^2$, [susy] & $-0.12500$ & $-0.11133$ & $+0.02265$ & $-0.25000$ & $-0.16016$ & $+0.04219$ \\
$SO(6)$ $S^2$, [susy] & $-0.10417$ & $-0.10295$ & $-0.00261$ & $-0.20833$ & $-0.15381$ & $-0.01061$ \\
$SO(3)$ $S^2$--$A^2$, [ns$_+$] & $-0.13158$ & $-0.10967$ & $+0.05226$ & $-0.14286$ & $-0.16760$ & $+0.04895$ \\ \midrule
large $N$ & $-0.10000$ & $-0.10100$ & $-0.01043$ & $-0.20000$ & $-0.15200$ & $-0.02694$ \\
\bottomrule
\end{tabular}
\caption{Coefficients of $\epsilon$, $\epsilon^2$, $\epsilon^3$ in $C_T/C_{T,\rm free}-1$ and $C_J/C_{J,\rm free}-1$ at fixed points of the matrix Gross--Neveu--Yukawa
models of \cite{PrakashSinha}, in their notation for the fixed points.  $C_J$ is for a generator of $SU(N)$ or $SO(N)$.  The large-$N$ row is \eqref{eq:matrixN}.
The $\epsilon^3$ terms include the odd fermion loops \eqref{eq:oddT} and \eqref{eq:oddJ}.}
\label{tab:matrix}
\end{table}

At large $N$ we evaluate the invariants in the double-line expansion, which gives the planar limit in closed form,
\begin{align}
\frac{C_T}{C_{T,\rm free}}&=1-\frac{\epsilon}{10}-\frac{101\,\epsilon^2}{1000}+\Big(\frac{13\,\zeta_3}{250}-\frac{21881}{300000}\Big)\epsilon^3+\mathcal O(\epsilon^4),\nonumber\\
\frac{C_J}{C_{J,\rm free}}&=1-\frac{\epsilon}{5}-\frac{19\,\epsilon^2}{125}+\Big(\frac{\zeta_3}{20}-\frac{272}{3125}\Big)\epsilon^3+\mathcal O(\epsilon^4).\label{eq:matrixN}
\end{align}
By the large-$N$ equivalence these hold for all three models, and the finite-$N$ values in table~\ref{tab:matrix} approach them.  None of these theories
has a four-dimensional counterpart, and the odd fermion loops \eqref{eq:oddT} and \eqref{eq:oddJ} survive the planar limit, where they
contribute $(\frac76-\zeta_3)\frac{\epsilon^3}{1250}$ to $C_T/C_{T,\rm free}$ and $\frac{\epsilon^3}{500}$ to $C_J/C_{J,\rm free}$.  For $C_T$ they are
small, but for $C_J$ they are as large as the $\epsilon^3$ coefficients. At the supersymmetric points of the $SO(3)$ and $SU(3)$ models they change
them from $-0.094$ and $-0.189$ to $+0.025$ and $-0.061$.

\subsection{Estimates in \texorpdfstring{$d=3$}{d=3}}\label{sec:susyd3}
\begin{figure}[htbp]
\centering
\includegraphics[width=0.95\textwidth]{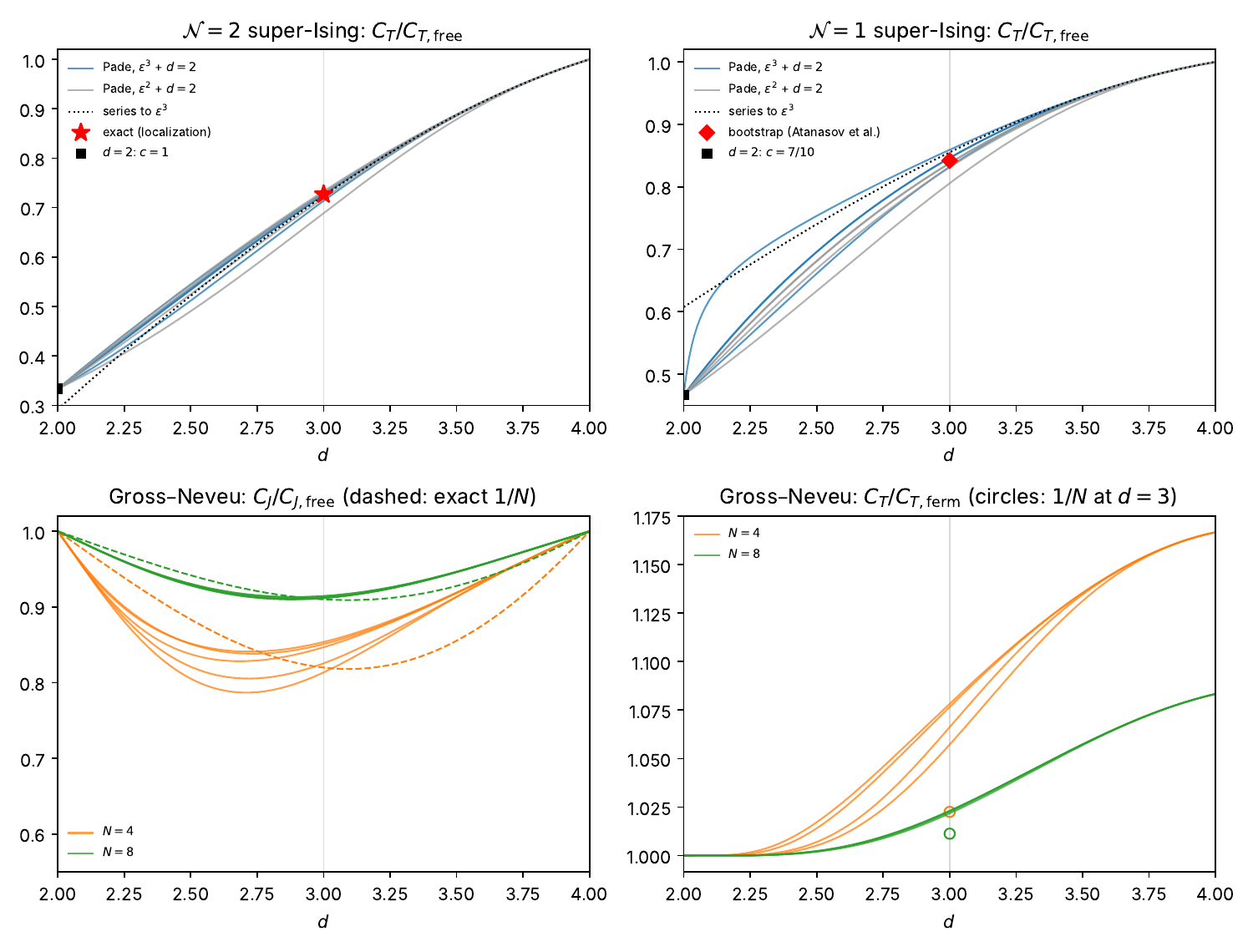}
\caption{Pole-free two-sided Pad\'e approximants for $2\le d\le4$ (sections~\ref{sec:susyd3} and~\ref{sec:d3}).  Top: $C_T/C_{T,\rm free}$ at the $\mathcal N=2$ (left) and
$\mathcal N=1$ (right) super-Ising points, with $\mathcal O(\epsilon^3)$ (blue) and $\mathcal O(\epsilon^2)$ (grey) input at $d=4$ and the exact $d=2$ value
(square); the dotted line is the truncated $\epsilon^3$ series, and the red symbol the exact \cite{NishiokaYonekura} or bootstrap \cite{Atanasov} value in $d=3$.
Bottom: Gross--Neveu $C_J/C_{J,\rm free}$ (left) and $C_T/C_{T,\rm ferm}$ (right) at $N=4$ and $N=8$, using the $2+\epsilon$ input of \cite{Diab}; dashed
curves and circles show the leading large-$N$ result.}
\label{fig:pade}
\end{figure}

To estimate $C_T$ and $C_J$ in $d=3$ we combine the $4-\epsilon$ series with information at $d=2$ in two-sided Pad\'e approximants, as in \cite{FGKT}
(see \cite{IhrigMihailaScherer} for critical exponents), who used $\mathcal O(\epsilon)$ input and noted that higher orders should improve the agreement.  An
approximant $[m,n]$ is a ratio of polynomials in $\epsilon$ of degrees $m$ and $n$ that reproduces the $4-\epsilon$ series through $\epsilon^K$ and the $d=2$
data, either a single value or a series in $d-2$ through order $L$; this fixes $m+n=K+L+1$.  Where no $d=2$ data are available we use the $4-\epsilon$ series
alone, with $m+n=K$.  We keep only approximants without poles for $2\le d\le4$, and quote the range they span.  Table~\ref{tab:d3} and figure~\ref{fig:pade}
collect the results, including those for the Gross--Neveu model of section~\ref{sec:d3}.

\paragraph{The super-Ising points.}
For the two supersymmetric points of \eqref{eq:susy} the $d=2$ end point is known exactly: the $\mathcal N=2$ minimal
model ($c=1$) and the tricritical Ising model ($c=7/10$), so that $C_T/C_{T,\rm free}=\frac13$ and $\frac7{15}$ at $d=2$.

For the $\mathcal N=2$ point, the truncated series at $\epsilon=1$ gives $0.7240$.  The pole-free approximants with $\mathcal O(\epsilon^3)$ input and the
$d=2$ value span $0.7139$--$0.7304$ (with $\mathcal O(\epsilon^2)$ input, $0.6890$--$0.7333$) and enclose the exact value $0.7268$. 

For the $\mathcal N=1$ point the approximants span $0.832$--$0.859$ and contain the bootstrap value $0.84207(7)$ \cite{Atanasov}.  
Our results are therefore consistent with the identification of the $\mathcal N=1$ fixed point of the chiral Ising model with the $\mathcal N=1$
superconformal theory found by the bootstrap.

\paragraph{The Liendo--Rong points.}
For the points of section~\ref{sec:LR} the $d=2$ theories are not known, so we use the $4-\epsilon$ series alone.  The pole-free Pad\'e approximants
give $0.820$--$0.830$ for the $S_4$ super-Potts point, $0.830$--$0.838$ for $S_5$ and $0.563$--$0.638$ for $O(3)/\mathbb Z_2$.

\paragraph{The matrix models.}
For the matrix models of section~\ref{sec:matrix} we again use the $4-\epsilon$ series alone.  At large $N$ the pole-free approximants give
$C_T/C_{T,\rm free}=0.787$--$0.805$ and $C_J/C_{J,\rm free}=0.615$--$0.672$ in $d=3$.  The supersymmetric points of table~\ref{tab:matrix} with $N=4$ and $6$ give
$C_T/C_{T,\rm free}$ between $0.765$ and $0.804$, close to the large-$N$ range, and the stable $S^2$--$A^2$ point at $N=3$ gives $0.794$--$0.811$.  For $N=3$ the
supersymmetric points have large coefficients, and their approximants spread widely.  For $C_J$ at the $SO(3)$ point, the $O(3)/\mathbb Z_2$ point,
the odd fermion loops matter: with them the approximants give $0.357$--$0.466$ (without them, $0.155$--$0.441$).

The supersymmetric point of the $SU(3)$ model, with superpotential proportional to $\det M$, is conjectured to be dual to $\mathcal N=2$ SQED with
two chiral flavors of charge $+1$, whose $SU(2)\times U(1)$ symmetry enhances to $SU(3)$ \cite{BeniniBenvenuti}.  For this theory localization
gives $C_T/C_{T,\rm free}=0.507$ and $C_J/C_{J,\rm free}=0.288$ \cite{GangYamazaki}, where the free values are those of four chiral multiplets and,
for $C_J$, of two with unit charge.  Our approximants give $0.36$--$0.56$ and $0.099$--$0.376$, which contain both values, although both ranges are
wide.

\begin{table}[htbp]
\centering\small
\begin{tabular}{>{\raggedright\arraybackslash}p{3.9cm}>{\raggedright\arraybackslash}p{5.9cm}>{\raggedright\arraybackslash}p{4.3cm}}\toprule
$d=3$ & this work (pole-free Pad\'e range) & comparison\\\midrule
$\mathcal N=2$ super-Ising, $C_T/C_{T,\rm free}$ & $0.7139$--$0.7304$; $[3,1]$ $0.7262$, $[4,0]$ $0.7266$, $[2,2]$ $0.7277$ & $0.7268$ exact \cite{NishiokaYonekura,WitczakKrempa,Bobev}; $0.77$ \cite{FGKT}\\
$\mathcal N=1$ super-Ising, $C_T/C_{T,\rm free}$ & $0.832$--$0.859$; $[4,0]$, $[2,2]$ $0.846$ & $0.84207(7)$ bootstrap \cite{Atanasov}; $0.86$ \cite{FGKT}\\
GN $N=8$, $C_J/C_{J,\rm free}$ & $0.912$--$0.914$ & $0.910$ large $N$; $0.902$ \cite{Diab}\\
GN $N=8$, $C_T/C_{T,\rm ferm}$ & $1.022$--$1.023$ & $1.011$ large $N$; $1.011$ \cite{Diab}\\
GN $N=4$, $C_T/C_{T,\rm ferm}$ & $1.057$--$1.078$ & $1.023$ large $N$; $1.015$ \cite{Diab}\\
chiral XY $N=8$, $C_T/C_{T,\rm ferm}$ & $1.025$--$1.073$; $[4,0]$, $[3,1]$ $1.025$ & $1.023$ large $N$ \cite{FGKT}\\\bottomrule
\end{tabular}
\caption{Estimates in $d=3$.  Ranges are spanned by the two-sided Pad\'e approximants $[m,n]$ (section~\ref{sec:susyd3}) that have no poles for $2\le d\le4$;
individual approximants are quoted where the range is wide.  $C_{T,\rm ferm}$ is the free-fermion part of $C_{T,\rm free}$.}
\label{tab:d3}
\end{table}

\section{Applications}\label{sec:applications}
\subsection{Gross--Neveu estimates in \texorpdfstring{$d=3$}{d=3}}\label{sec:d3}
For the Gross--Neveu (chiral Ising) model we use the two-sided Pad\'e approximants of section~\ref{sec:susyd3}, and combine our series with the $2+\epsilon$ expansions of \cite{Diab} ($C_J$ to $\epsilon^2$, $C_T$ to $\epsilon^3$, the latter through their function $f(d)$).
The same procedure reproduces the exact large-$N$ coefficients in $d=3$.  At $N=400$ the approximants give $N(C_J/C_{J,\rm free}-1)\in[-0.727,-0.721]$ and $N(C_T/C_{T,\rm ferm}-1)\in[0.086,0.099]$,
against $-64/(9\pi^2)=-0.721$ and $8/(9\pi^2)=0.090$.  At smaller $N$ the $C_T$ approximants lie well above the leading large-$N$ value.  Empirically $N(C_T/C_{T,\rm ferm}-1)\approx0.090+0.69/N$, and at $N=8$
the approximants give $0.174$--$0.184$, about twice the $1/N$ coefficient; taken at face value, this means a large $1/N^2$ correction.
At $N=4$, however, the pole-free range with $\mathcal O(\epsilon^3)$ input, $1.057$--$1.078$, does not overlap the range $1.0147$--$1.0156$ obtained with the
$\mathcal O(\epsilon)$ input of \cite{Diab}.  The spread between input orders, rather than the spread among approximants of one order, is therefore the more realistic
measure of the uncertainty, and an independent determination at small $N$ would be needed to establish the size of the $1/N^2$ term.

\subsection{Universal conductivity at Dirac quantum critical points}
In $2+1$ dimensions the zero-temperature conductivity of a conformal field theory is fixed by $C_J$ of the corresponding current
\cite{HuhStrackSachdev,HuhStrack,WitczakKrempa,Diab}, so that $\sigma/\sigma_{\rm free}=C_J/C_{J,\rm free}$.  The best-known setting for such critical points is graphene \cite{CastroNeto}.  Its electrons hop on a honeycomb lattice, whose two
sublattices play the role of the spinor index; near the two inequivalent corners (``valleys'') of the Brillouin zone the low-energy excitations are
massless two-component Dirac fermions in $2+1$ dimensions, and two valleys and two spin components give $N=8$ in our normalization.  Short-range
interactions are irrelevant at the free fixed point, but sufficiently strong repulsion drives a transition into an ordered state, whose order
parameter $\phi$ gaps the fermions through $\phi\,\bar\psi M\psi$ \cite{Herbut2006,HerbutJuricicVafek,Zerf}.  A charge-density wave, in which one
sublattice is more occupied than the other, has a real order parameter and is in the chiral Ising class; a Kekul\'e bond-order state has a complex
order parameter and is in the chiral XY class; and a N\'eel antiferromagnet has a three-component order parameter and is in the chiral Heisenberg
class \cite{HerbutJuricicVafek,RoyJuricicHerbut,Zerf}, which has been studied in quantum Monte Carlo simulations of the Hubbard model
\cite{SorellaTosatti,AssaadHerbut}.  For chiral Ising and XY, $N=4$ describes spinless fermions on this lattice.

In all three cases the
order parameter is electrically neutral, so the electric current is the fermion number, which acts on the fermions only.  For graphene,
$\sigma_{\rm free}=\pi e^2/(2h)$ is the conductivity of free Dirac fermions.  The long-range Coulomb interaction is irrelevant at these critical points \cite{HerbutJuricicVafek}.  The conductivity at such critical points is universal, but its value has been left open \cite{RoyJuricicHerbut}.

For the chiral Ising model $C_J/C_{J,\rm free}$ is given by \eqref{eq:isingCJ}, and our results give its $\epsilon^3$ term.  For chiral XY and Heisenberg we evaluate
\eqref{eq:CJ} and \eqref{eq:CJ3} for the fermion-number current, which has $\theta=0$.  Its generator commutes with every $y^a$ and squares to $\mathbf 1$ on the fermions,
so each invariant takes the same value as for $\Theta=\mathbf 1$.  We evaluate these invariants at the fixed points of section~\ref{sec:models}, and table~\ref{tab:sigma} gives the resulting series.
At large $N$ the leading correction is $n_s$ times the Gross--Neveu value.  Our series confirm $C_{J1}=n_s\,C_{J1}^{\rm GN}$ through
$\epsilon^3$, and in $d=3$ the exact result is $1-64n_s/(9\pi^2N)$.

For $d=3$ we form two-sided Pad\'e approximants that match the $\epsilon^3$ series and the free value $C_J/C_{J,\rm free}=1$ at $d=2$.  The latter assumes
that, as for the Gross--Neveu model, the fixed points connect to free fermions as $d\to2$, where the four-fermion description in $d=2+\epsilon$ is
asymptotically free; it holds exactly at large $N$, where $C_{J1}\propto d-2$.  Two tests support the procedure.  For chiral Ising, where the $2+\epsilon$
expansion of $C_J$ is known \cite{Diab}, the approximants with only the $d=2$ value give $0.904$--$0.927$ at $N=8$, which contains the range
$0.912$--$0.914$ obtained with the full $2+\epsilon$ input (section~\ref{sec:d3}).  At $N=400$ they give $N(C_J/C_{J,\rm free}-1)$ within $12\%$
of the exact $-64n_s/(9\pi^2)$ for all three models, enclosing it in each case.

\begin{table}[h]
\centering\small\setlength{\tabcolsep}{3pt}
\begin{tabular}{lccccccc}
\toprule
& \multicolumn{3}{c}{$N=8$ coefficients} & \multicolumn{2}{c}{$N=8$, $d=3$} & \multicolumn{2}{c}{$N=4$, $d=3$}\\
model (ordered state) & $\epsilon$ & $\epsilon^2$ & $\epsilon^3$ & Pad\'e & large $N$ & Pad\'e & large $N$\\ \midrule
chiral Ising (CDW) & $-0.1071$ & $-0.0193$ & $+0.0314$ & $0.912$--$0.914$ & $0.910$ & $0.814$--$0.854$ & $0.820$\\
chiral XY (Kekul\'e) & $-0.2500$ & $-0.0217$ & $+0.0694$ & $0.785$--$0.831$ & $0.820$ & $0.645$--$0.686$ & $0.640$\\
chiral Heisenberg (N\'eel) & $-0.4500$ & $-0.0111$ & $+0.1665$ & $0.640$--$0.738$ & $0.730$ & $0.428$--$0.489$ & $0.460$\\
\bottomrule
\end{tabular}
\caption{$\sigma/\sigma_{\rm free}=C_J/C_{J,\rm free}$ of the electric (fermion-number) current.  Pad\'e ranges are over pole-free two-sided approximants
with the $\epsilon^3$ series and the $d=2$ value; for chiral Ising they use the full $2+\epsilon$ input of \cite{Diab}.  Large $N$: $1-64n_s/(9\pi^2N)$.}
\label{tab:sigma}
\end{table}

In all the cases we have examined, the estimates lie close to the leading large-$N$ value, and enclose it in most cases, even at $N=4$.  Unlike $C_T$, the conductivity shows
no sign of a large $1/N^2$ correction.  Therefore we conclude that, for graphene, $\sigma/\sigma_{\rm free}\approx0.91$, $0.79$--$0.83$ and $0.64$--$0.74$ at the
charge-density-wave, Kekul\'e and N\'eel transitions.

For the semimetal--superconductor transition on the surface of a topological insulator \cite{GroverShengVishwanath,PonteLee,RoyJuricicHerbut,ZerfLinMaciejko,WitczakKrempa},
the conductivity is known exactly.  A single Dirac cone couples to the charge-2 Cooper-pair field through $\phi^*\psi^Ti\gamma_2\psi+{\rm c.c.}$, so the critical point is the chiral XY model
with $N=2$, and the electric current is its $U(1)$ current, with charge 1 on the fermion and 2 on the boson.  This fixed point has emergent $\mathcal N=2$ supersymmetry, and
the electric current is three times the superconformal $R$-current there, but not at the free fixed point (section~\ref{sec:Rcurrent}).  By \eqref{eq:TIfactor}, with the exact $C_T$ from localization \cite{NishiokaYonekura,WitczakKrempa},
$C_T/C_{T,\rm free}=4\tau_{RR}=\frac{16}{243}\big(16-9\sqrt3/\pi\big)$,
\begin{equation}\label{eq:TIsigma}
\frac{\sigma}{\sigma_{\rm free}}=\frac{9}{10}\,\frac{C_T}{C_{T,\rm free}}=\frac{8\,(16\pi-9\sqrt3)}{135\,\pi}\approx0.654,
\end{equation}
where $\sigma_{\rm free}$ is the conductivity of a free Dirac fermion and a free charge-2 boson.  In units of $e^2/\hbar$ this is
$\sigma=(16\pi-9\sqrt3)/(54\pi)=\frac98\tau_{RR}\approx0.204$.\footnote{Reference~\cite{WitczakKrempa} fixed the constant relating the electric current to the $R$-current at the free fixed point, where
$C_J^{\rm el}/C_T=\frac53$, and obtained $\sigma/\sigma_{\rm free}=C_T/C_{T,\rm free}\approx0.727$.  Since the two currents are proportional only at the
interacting fixed point, where $C_J^{\rm el}/C_T=\frac32$, the conductivity is $\frac9{10}$ of that value.  The shear viscosity and the corner coefficient
of the entanglement entropy depend only on $C_T$ and are unaffected, as is the result $\gamma=0$ for $\langle TJJ\rangle$, but the universal ratios become
$\sigma_\infty/\eta_\infty=36$ instead of $40$ and $\lambda_{\rm corner}=\sigma_\infty/18$ instead of $\sigma_\infty/20$.}

\section{Conclusion}\label{sec:conclusion}
We have computed $C_T$ and $C_J$ through $\mathcal O(\epsilon^3)$ at any fixed point of the general renormalizable scalar--fermion theory in
$d=4-\epsilon$, as a small number of coupling-tensor invariants with $\epsilon$-independent coefficients.  The three-loop Yukawa coefficients and
all $\mathcal O(\epsilon^3)$ terms are new, to the best of our knowledge.  The results include three-dimensional theories without a
four-dimensional counterpart, where fermion loops with an odd number of Yukawa vertices contribute; we fixed their continuation away from $d=3$ with
$\mathcal N=1$ supersymmetry.  In $d=3$ the results agree well with the exact $C_T$ of the $\mathcal N=2$ super-Ising model and with the bootstrap
value at the $\mathcal N=1$ point, they are consistent with the exact $C_T$ and $C_J$ of $\mathcal N=2$ SQED with two flavors at its
conjectured dual, the $SU(3)$ matrix model, and they give the universal conductivity at the charge-density-wave, Kekul\'e and N\'eel transitions of
graphene.  At the supersymmetric transition on the surface of a topological insulator, the electric current is three times the superconformal
$R$-current only at the interacting fixed point, which gives the exact conductivity $\sigma/\sigma_{\rm free}=\frac9{10}\,C_T/C_{T,\rm free}\approx0.654$.

Our results have a few limitations.  Firstly, most individual $\mathcal O(\epsilon^3)$ coefficients largely rely only on computation via Forcer, since large $N$, the $O(n)$ model and supersymmetry, with four supercharges and, through the supercurrent, with two, together constrain only a few linear combinations of them.  Secondly, the
continuation of the odd fermion loops is fixed by supersymmetry at the supersymmetric points; at the non-supersymmetric three-dimensional-only points, chiral Ising at $N=2$ with its $U(1)$ symmetry and the $SO(3)$ $S^2$--$A^2$ [ns$_+$] point of the
matrix models, we use the same continuation, assuming that it is a property of the scheme rather than of the theory.  Why supersymmetry selects this
continuation, and whether it can be derived within dimensional regularization, remains to be understood; we will return to this question, using a
formulation of the continuation in terms of Deligne categories, in a separate paper.
Thirdly, for the Gross--Neveu model at small $N$ the $d=3$ estimates of $C_T$ depend noticeably on the order of the input, so the size of the $1/N^2$ correction remains uncertain.

We close with a few directions for further work.  The most natural extension is to gauge--Yukawa theories, such as the QED$_3$--Gross--Neveu--Yukawa
models of Dirac spin liquids, where the same method applies with gauge-field propagators and the gauge contributions to the $\beta$-functions.  Finally, it would be interesting to compute the conformal-collider parameters \cite{HofmanMaldacena}.

\section*{Code and data availability}\label{sec:data}
Code and data  are available at the GitHub repository \href{https://github.com/shubham-kumar-sinha/yukawa-ctcj}{\texttt{yukawa-ctcj}}~\cite{yukawactcj}.  The repository contains every diagram class through four loops with its graph and weight, the full $\mathcal O(\epsilon^3)$ coefficients in machine-readable form, and the code that reproduces every table, figure and check
of this paper, including the supercurrent and flavor-current computation of appendix~\ref{app:supercurrent}.  With FORM~5.0.2 \cite{FORM5} and
Forcer \cite{Forcer} installed, every integral can be recomputed.

\section*{Acknowledgments}
This work was supported in part by the DST-SERB grant (CRG/2021/009137). SS thanks Diptarka Das for explaining aspects of $C_T$ and $C_J$, and for supporting an extended stay at IIT Kanpur in May--June 2024, where part of this work was completed.  We acknowledge the use of a large-language model (Claude Opus~5.5 and Claude Fable~5.2) for assistance in writing code and drafting the text; the authors take full responsibility for the content of this paper.

\appendix
\section{Feynman rules }\label{app:method}

\paragraph{Fourier transforms and propagators.}
We use $\int_p=\int d^Dp/(2\pi)^D$, and $\int_{p,q}=\int_p\int_q$. Here, $D=4-\epsilon=4-2\epsilon'$ (so $D=d$; we write $D$ where we follow the Forcer conventions). The Fourier transforms of the fields are
\begin{equation}\label{eq:app-field-fourier}
\begin{aligned}
\phi_a(x)&=\int_q e^{iq\cdot x}\phi_a(q),\qquad
\psi_j(x)=\int_{p_{\rm in}}e^{ip_{\rm in}\cdot x}\psi_j(p_{\rm in}),\\
\bar\psi_i(x)&=\int_{p_{\rm out}}e^{-ip_{\rm out}\cdot x}\bar\psi_i(p_{\rm out}).
\end{aligned}
\end{equation}
Overall momentum-conserving delta functions are suppressed below.  A field momentum $q$ in
$e^{iq\cdot x}$ flows into the vertex at $x$.  For the composite operators we define $O(Q)$ by $O(x)=\int_Q e^{-iQ\cdot x}O(Q)$, so that $Q$ enters
the diagram at $x$.  The two-point functions are even in $Q$, so the sign of the exponent in the inverse
transform \eqref{eq:app-scalar-fourier} is immaterial.
The scalar and fermion propagators, with spinor indices suppressed, are
\begin{equation}\label{eq:app-propagators}
\Delta_{ab}(p)=\frac{\delta_{ab}}{p^2},\qquad
S_{ij}(p)=-i\,\delta_{ij}\frac{p\!\!\!/}{p^2}.
\end{equation}

\paragraph{Interaction vertices.}
We denote the momentum-space vertex factor by $K$, with the field labels fixed. The combinatorial factors are included in the class weight of section~\ref{sec:diagrams}. In the conventions of \eqref{eq:action}, we have:
\begin{equation}\label{eq:app-interactions}
K_{\phi_a\bar\psi_i\psi_j}=-(y^a)_{ij},\qquad
K_{\phi_a\phi_b\phi_c\phi_d}=-\lambda_{abcd}.
\end{equation}
Their minus signs come from the expansion of $e^{-\mathcal S_{\rm int}}$.
We take fermion momenta along the arrows. Each closed fermion loop contributes a factor $-1$, with the matrices in the spinor and flavor traces ordered against the fermion arrows.  Integrating out Majorana fermions gives a Pfaffian, $(\det)^{1/2}$, of the operator whose determinant a Dirac fermion would give, so each closed loop carries a factor $\frac12$ relative to these Dirac-type rules.  This factor is included in $\Tr$.

\paragraph{Stress-tensor insertions.}
The Lorentz indices are left open at each insertion and contracted with the transverse-traceless projector in the full diagram.
The kinetic parts of \eqref{eq:TJ} are
\begin{equation}\label{eq:app-T-operators}
(T_\phi)_{\mu\nu}=\partial_\mu\phi_a\,\partial_\nu\phi_a,\qquad
(T_\psi)_{\mu\nu}=\frac12\bar\psi\,\gamma_{(\mu}\overleftrightarrow\partial_{\nu)}\psi,
\end{equation}

For scalar legs with momenta $q_1$, $q_2$ flowing into the vertex, and fermion momenta $p_{\rm in}$, $p_{\rm out}$, $\overleftrightarrow\partial_\nu$ gives
$i(p_{\rm in}+p_{\rm out})_\nu$.  Since the projector is symmetric in each index pair, we may use the unsymmetrized insertion factors
\begin{equation}\label{eq:app-T-vertices}
\begin{aligned}
K_{T_\phi,\mu\nu}^{ab}(q_1,q_2)&=-\delta_{ab}(q_1)_\mu(q_2)_\nu,\\
K_{T_\psi,\mu\nu}^{ij}(p_{\rm out},p_{\rm in})&=\frac i2\,\delta_{ij}\,\gamma_\mu(p_{\rm in}+p_{\rm out})_\nu.
\end{aligned}
\end{equation}

The $\xi$-term and all terms proportional to $\delta_{\mu\nu}$, including the interaction terms of the stress tensor, vanish under the transverse-traceless projection.

\paragraph{Current insertions.}
For the current convention of \eqref{eq:TJ}, namely the Noether current divided by $i$, the fermion and scalar parts are
\begin{equation}\label{eq:app-J-operators}
(J_\psi)_\mu=\bar\psi\gamma_\mu \Theta\,\psi,\qquad
(J_\phi)_\mu=-i(\partial_\mu\phi_a)\theta_{ab}\phi_b.
\end{equation}
The insertion factors are
\begin{equation}\label{eq:app-J-vertices}
K_{J_\phi,\mu}^{ab}(q_1,q_2)=\theta_{ab}(q_1)_\mu,\qquad
K_{J_\psi,\mu}^{ij}(p_{\rm out},p_{\rm in})=\gamma_\mu \Theta_{ij}.
\end{equation}
All insertions are written for fixed assignments of the field legs, whose Wick contractions are included in the class weight.  The two current indices are contracted with the transverse projector in the full diagram.

\paragraph{Projections and Feynman rules.}
We contract the current and stress-tensor indices with the transverse
and transverse-traceless projectors given in \eqref{eq:projector}.

Let $G_T^{\mu\nu,\rho\sigma}(Q)$ and $G_J^{\mu\nu}(Q)$ denote the momentum-space two-point functions with the overall delta function removed. We compute
\begin{equation}\label{eq:app-projected-scalars}
\mathcal R_T=(D-1)P_{\mu\nu,\rho\sigma}G_T^{\mu\nu,\rho\sigma},\qquad
\mathcal R_J=\pi_{\mu\nu}G_J^{\mu\nu}.
\end{equation}
Using $P^2=P$, $\pi^2=\pi$ and
\begin{equation}\label{eq:app-projector-traces}
P_{\mu\nu,\mu\nu}=\frac{(D+1)(D-2)}{2},\qquad \pi_{\mu\mu}=D-1,
\end{equation}
we obtain the form factors in \eqref{eq:decomp}:
\begin{equation}\label{eq:app-projection-normalization}
A_T(Q^2)=\frac{2\mathcal R_T}{(D-1)(D+1)(D-2)(Q^2)^2},\qquad
A_J(Q^2)=\frac{\mathcal R_J}{(D-1)Q^2}.
\end{equation}
The contribution of diagram class $\mathcal{G}$ with $L$ loops to $\mathcal{R}_O$, $O=T,J$ is a flavor factor $F_{\mathcal{G}}$ times $\mathcal{R}_{O,\mathcal{G}}$ which carries no flavor indices.   The flavor factor collects the couplings $y^a$ and $\lambda_{abcd}$, the generators $\Theta$ and $\theta$, the Kronecker deltas of the propagators, and one flavor trace for each fermion loop.  Everything else belongs to $\mathcal R_{O,\mathcal G}$: the
weight $w_{\mathcal G}$, the vertex signs, the momentum and spinor parts of the propagators and vertex factors, and the sign of each fermion loop. For a closed fermion loop the split is fixed by the trace normalization of section~\ref{sec:theory}.  The loop contributes $-\Tr$ of the
product of its propagators, vertices and insertions, ordered against the fermion arrows.  Each factor is a spinor matrix times a flavor matrix, so the
product is $\Sigma\otimes M$, where $\Sigma$ is the ordered product of $\gamma$-matrices and $M$ the corresponding product of $y^a$ and $\Theta$.  The trace
then factorizes as $\Tr(\Sigma\otimes M)=\frac12\mathrm{tr}(\Sigma)\,\hat{\Tr}(M)$: the $\frac12$ is the Majorana factor, the spinor trace is normalized to $\mathrm{tr}\,\mathbf 1=2$, so that
$\frac12\mathrm{tr}\,\mathbf 1=1$, and $\hat{\Tr}(M)$ is the flavor trace, with $\hat{\Tr}\mathbf 1=N$.  The spinor trace $\mathrm{tr}(\Sigma)$ belongs to $\mathcal R_{O,\mathcal G}$ and the flavor trace $\hat{\Tr}(M)$ to $F_{\mathcal G}$.

For each diagram class, FORM builds the integrand of $\mathcal{R}_{O,\mathcal{G}}$ from the propagators and vertex factors, with the momenta routed through the graph.  It contracts the open indices with the projectors \eqref{eq:projector}, evaluates the spinor traces and sets $Q^2=1$.  The result is a sum of $L$-loop propagator integrals with scalar-product numerators.  Forcer reduces them to master integrals and expands the result in $\epsilon'=\epsilon/2$ in the G-scheme, in which each loop is counted in units of $\epsilon'G(1,1)$, where
\begin{equation}\label{eq:app-G11}
\int_p\frac{1}{p^2(p+Q)^2}=G(1,1)\,(Q^2)^{-\epsilon'},\qquad
G(1,1)=\frac{\Gamma(1-\epsilon')^2\,\Gamma(\epsilon')}{(4\pi)^{2-\epsilon'}\,\Gamma(2-2\epsilon')} .
\end{equation}
The output for the class is the Laurent series
\begin{equation}\label{eq:app-forcer-output}
X_{O,\mathcal G,L}(\epsilon')=\frac{\mathcal R_{O,\mathcal G}(Q)\big|_{Q^2=1}}{\big[\epsilon'G(1,1)\big]^L}.
\end{equation}
The value at
$Q^2=1$ suffices: rescaling all loop momenta by $(Q^2)^{1/2}$ shows that the $L$-loop part of $\mathcal R_T$ is $(Q^2)^{2-L\epsilon'}$ times its value at
$Q^2=1$, and that of $\mathcal R_J$ is $(Q^2)^{1-L\epsilon'}$ times its value at $Q^2=1$.

The flavor factors are invariants, which do not change under orthogonal changes of the scalar and fermion bases.  We expand each of them in a basis
$\{I_k\}$ of independent invariants,
\begin{equation*}
F_{\mathcal G}=\sum_k d_{\mathcal G k}\,I_k,
\end{equation*}
with rational coefficients $d_{\mathcal G k}$. The $L$-loop G-scheme value of $I_k$ is then
\begin{equation}\label{eq:app-invariant-sum}
X_{O,k,L}(\epsilon')=\sum_{\mathcal G}d_{\mathcal G k}\,X_{O,\mathcal G,L}(\epsilon'),
\end{equation}
where the sum runs over the classes with $L$ loops.  By \eqref{eq:app-projection-normalization}--\eqref{eq:app-invariant-sum}, the $L$-loop part of each
form factor is a single power of $Q^2$,
\begin{equation}\label{eq:app-gscheme-form-factors}
\begin{aligned}
A_O^{(L)}(Q^2)&=a_{O,L}\,(Q^2)^{-L\epsilon'},\\
a_{T,L}&=\big[\epsilon'G(1,1)\big]^L\sum_k\frac{2X_{T,k,L}}{(D-1)(D+1)(D-2)}\,I_k,\\
a_{J,L}&=\big[\epsilon'G(1,1)\big]^L\sum_k\frac{X_{J,k,L}}{D-1}\,I_k,
\end{aligned}
\end{equation}
and $A_O=\sum_LA_O^{(L)}$.
At one loop, with the class weights included (2 for the $T_\phi T_\phi$ diagram, from the two pairings of its legs, and 1 for the others), the diagrams give, exactly in $D$,
\begin{equation}\label{eq:app-one-loop-X}
\begin{aligned}
X_{T,\Tr\mathbf 1,1}&=\frac{(D-1)(D-2)}{16\,\epsilon'},&\qquad X_{T,n_s,1}&=\frac{D-2}{8\,\epsilon'},\\
X_{J,\Tr(\Theta \Theta),1}&=\frac{D-2}{2\,\epsilon'},&\qquad X_{J,\theta_{ab}\theta_{ab},1}&=\frac{1}{2\,\epsilon'}.
\end{aligned}
\end{equation}
With \eqref{eq:app-gscheme-form-factors} and the Fourier transform below, these values reproduce the free
values \eqref{eq:free} exactly in $d$.

\paragraph{Fourier transform and extraction of $C_T$ and $C_J$.}

Recall that in position space, $C_T$ and $C_J$ are the normalizations of the conformal two-point functions
\begin{equation}\label{eq:app-conformal}
\langle T_{\mu\nu}(x)\,T_{\rho\sigma}(0)\rangle=C_T\,\frac{I_{\mu\nu,\rho\sigma}(x)}{x^{2d}},\qquad
\langle J_\mu(x)\,J_\nu(0)\rangle=-C_J\,\frac{I_{\mu\nu}(x)}{x^{2(d-1)}},
\end{equation}
where the minus sign holds because $C_J$ is defined by the Noether current $iJ$ (section~\ref{sec:method}), $I_{\mu\nu}=\delta_{\mu\nu}-2x_\mu x_\nu/x^2$ and
$I_{\mu\nu,\rho\sigma}=\frac12(I_{\mu\rho}I_{\nu\sigma}+I_{\mu\sigma}I_{\nu\rho})-\delta_{\mu\nu}\delta_{\rho\sigma}/d$.  A single tensor
structure suffices to read off each normalization.  We use the one with the most factors of $x$.  Only
$I_{\mu\rho}I_{\nu\sigma}$ and $I_{\mu\sigma}I_{\nu\rho}$ contain four of them, so
\begin{equation}\label{eq:app-target}
\langle T_{\mu\nu}(x)\,T_{\rho\sigma}(0)\rangle\supset 4C_T\,\frac{x_\mu x_\nu x_\rho x_\sigma}{(x^2)^{d+2}},\qquad
\langle J_\mu(x)\,J_\nu(0)\rangle\supset 2C_J\,\frac{x_\mu x_\nu}{(x^2)^{d}}.
\end{equation}

In momentum space, \eqref{eq:app-gscheme-form-factors} gives the $L$-loop part of each form factor as a pure power,
\begin{equation}\label{eq:app-loop-power}
A_O^{(L)}(Q^2)=a_{O,L}\,(Q^2)^{\kappa_L},\qquad \kappa_L=-\frac{L\epsilon}{2},\qquad O=T,J.
\end{equation}
For $x\neq0$, and up to contact terms, the Fourier transform of a power is
\begin{equation}\label{eq:app-scalar-fourier}
\int\frac{d^dQ}{(2\pi)^d}\,e^{iQ\cdot x}\,(Q^2)^\kappa=\frac{c(\kappa)}{(x^2)^{s}},\qquad s=\frac d2+\kappa,\qquad
c(\kappa)=\frac{4^\kappa\,\Gamma(d/2+\kappa)}{\pi^{d/2}\,\Gamma(-\kappa)} .
\end{equation}
We transform the spin-two part $Q^4P_{\mu\nu,\rho\sigma}A_T$ of \eqref{eq:decomp} and the current part $Q^2\pi_{\mu\nu}A_J$.  The spin-zero part
$Q^4\pi_{\mu\nu}\pi_{\rho\sigma}B_T$ also contains a term $Q_\mu Q_\nu Q_\rho Q_\sigma B_T$ without deltas, but it vanishes at the fixed point for $x\neq0$
(section~\ref{sec:ops}); away from the fixed point, \eqref{eq:app-CTCJx} below defines $C_T(x)$ through the spin-two part.  The momenta become derivatives, $Q_\mu\to-i\partial_\mu$.  In
$Q^4P_{\mu\nu,\rho\sigma}$ the only term without a Kronecker delta is $\frac{d-2}{d-1}Q_\mu Q_\nu Q_\rho Q_\sigma$, and in $Q^2\pi_{\mu\nu}$ it is $-Q_\mu Q_\nu$.
Every other term carries a delta and cannot produce the structures \eqref{eq:app-target}.  Since
\[
\partial_{\mu_1}\cdots\partial_{\mu_n}(x^2)^{-s}\supset(-2)^n\,(s)_n\,\frac{x_{\mu_1}\cdots x_{\mu_n}}{(x^2)^{s+n}},\qquad (s)_n=s(s+1)\cdots(s+n-1),
\]
the $L$-loop terms transform into
\begin{equation}\label{eq:app-fourier-tensor}
\begin{aligned}
Q^4P_{\mu\nu,\rho\sigma}A_T^{(L)}&\;\longrightarrow\;\frac{16(d-2)}{d-1}\,(s_L)_4\,c(\kappa_L)\,a_{T,L}\,
\frac{x_\mu x_\nu x_\rho x_\sigma}{(x^2)^{s_L+4}}+\dots,\\
Q^2\pi_{\mu\nu}A_J^{(L)}&\;\longrightarrow\;4\,(s_L)_2\,c(\kappa_L)\,a_{J,L}\,\frac{x_\mu x_\nu}{(x^2)^{s_L+2}}+\dots,
\end{aligned}
\end{equation}
with $s_L=d/2+\kappa_L$; the dots stand for terms containing $\delta$'s.  The exponents are $s_L+4=d+2-(L-1)\epsilon/2$ and $s_L+2=d-(L-1)\epsilon/2$.  Comparing with
\eqref{eq:app-target}, the $L$-loop term therefore contributes $b_{O,L}\,(x^2)^{(L-1)\epsilon/2}$ to $C_O$, with
\begin{equation}\label{eq:app-fourier-amplitudes}
b_{T,L}=\frac{4(d-2)}{d-1}\,(s_L)_4\,c(\kappa_L)\,a_{T,L},\qquad
b_{J,L}=2\,(s_L)_2\,c(\kappa_L)\,a_{J,L}.
\end{equation}

At one loop, $s_1=d-2$ and the power of $x^2$ is exactly conformal.  Then \eqref{eq:app-fourier-amplitudes} is the inverse of \eqref{eq:NTNJ},
$b_{O,1}=a_{O,1}/N_O$, and with \eqref{eq:app-one-loop-X} it gives the free values \eqref{eq:free}.  Beyond one loop the factor $(x^2)^{(L-1)\epsilon/2}$ is
balanced by the couplings.  The amplitude $a_{O,L}$ is built from the bare couplings $y_0$ and $\lambda_0$.  Their renormalization,
$y_0=\mu^{\epsilon/2}(y+\dots)$ and $\lambda_0=\mu^\epsilon(\lambda+\dots)$ (section~\ref{sec:method} and appendix~\ref{app:4loop}), turns each term into
$(\mu^2x^2)^{(L-1)\epsilon/2}$ times renormalized couplings and counterterms.  The sums
\begin{equation}\label{eq:app-CTCJx}
C_T(x)=\sum_L b_{T,L}\,(x^2)^{(L-1)\epsilon/2},\qquad C_J(x)=\sum_L b_{J,L}\,(x^2)^{(L-1)\epsilon/2}
\end{equation}
are then functions of the renormalized couplings and of $\log\mu^2x^2$.  Their poles in $\epsilon$ cancel separately in every invariant, and
$\log\mu^2x^2$ multiplies only terms proportional to the $\beta$-functions.  At a fixed point $C_T(x)$ and $C_J(x)$ are therefore independent of $x$, and
equal $C_T$ and $C_J$.

\section{Checks}\label{sec:checks}

In this appendix we collect the checks of our results against independent information; all of them pass.  We have also checked internal
consistency throughout. The free values are reproduced exactly in $d$, the poles cancel separately in every invariant, $\log\mu^2x^2$ drops out at
fixed points, and the Ward-identity relations among the $\langle JJ\rangle$ invariants hold.  At the end of appendix~\ref{app:chkCJ3} we comment on
the extent to which the checks constrain the individual $\mathcal O(\epsilon^3)$ coefficients.

\subsection{Through \texorpdfstring{$\mathcal O(\epsilon^2)$}{O(epsilon\^{}2)}}
\begin{itemize}
\item An earlier, unpublished computation of ours, with its own diagram generator, integrands, Fourier transform and renormalization but the same
FORM/Forcer back end, gives identical two-loop $\langle TT\rangle$ diagrams, large-$N$ family sum and finite-$N$ Gross--Neveu--Yukawa $C_T$ through
$\epsilon^2$, \eqref{eq:isingCT}.
\item The $\lambda^2$ terms reproduce the $O(n)$ results $-\frac{5(n+2)}{12(n+8)^2}\epsilon^2$ for $C_T$ \cite{OS} and $-\frac{3(n+2)}{4(n+8)^2}\epsilon^2$
for $C_J$ \cite{Petkou}.
\item The $\mathcal O(\epsilon)$ term of $C_T$ in the Gross--Neveu--Yukawa model agrees with \cite{FGKT}.
\item At large $N$, $C_{T1}=\frac23-\frac{11}{18}\epsilon-\frac{17}{54}\epsilon^2$ and $C_{J1}=-\frac32\epsilon+\frac12\epsilon^2$ agree with the exact
Gross--Neveu results of \cite{Diab}.
\item The supersymmetric relations hold: $C_J^R/C^R_{J,\rm free}=C_T/C_{T,\rm free}$ at the $\mathcal N=2$ points (section~\ref{sec:susychiral}), and
$C_S/C_{S,\rm free}=C_T/C_{T,\rm free}$ and $C_j/C_{j,\rm free}=C_J/C_{J,\rm free}$ at the $\mathcal N=1$ points (section~\ref{sec:oddsusy}).  These
use diagrams, projectors and counterterms independent of those of $\langle TT\rangle$, and they continue to hold at $\mathcal O(\epsilon^3)$.
\end{itemize}

\subsection{\texorpdfstring{$C_T$ at $\mathcal O(\epsilon^3)$}{C\_T at O(epsilon\^{}3)}}
Consider first the $O(n)$ model.  For $n=1,\dots,4$ our result gives $C_T/C_{T,\rm free}=1-\frac{5(n+2)}{12(n+8)^2}\epsilon^2-\frac{(n+2)(7n^2+382n+1708)}{36(n+8)^4}\epsilon^3$,
which for Ising is $-\frac{233}{8748}\epsilon^3$ at this order.  This is the analytic-bootstrap result \cite{GKSS,HvL}, which for $n=1$ is also reproduced by the direct four-loop
computation of \cite{IK}.  The $\lambda^3$ coefficient $-\frac{7}{36}$ is the four-loop coefficient $\alpha$ of appendix~E of \cite{OS}; given $-\frac5{36}$ and the two-loop $\lambda_*$,
it is already fixed by the Ising value.

At leading order in $1/N$, $\Tr(y_*^ay_*^a)\to\epsilon$ exactly and only the triple-trace $y^6$ structure contributes, giving
\[
C_{T1}=\frac23-\frac{11}{18}\epsilon-\frac{17}{54}\epsilon^2+\Big(\frac{\zeta_3}{2}-\frac{365}{2592}\Big)\epsilon^3+\mathcal O(\epsilon^4),
\]
which agrees with the exact Gross--Neveu result \cite{Diab} expanded to this order.

Large $N$ thus tests only the triple-trace coefficient, and the $O(n)$ model only the $\lambda$-only coefficients.  The $y^4\lambda$ and $y^2\lambda^2$
coefficients and the individual single- and double-trace $y^6$ coefficients come from the $\epsilon^{\prime-1}$ terms of the four-loop integrals, which pole cancellation
does not test.  They are tested by the expansion-depth check of section~\ref{sec:renorm} and by the $\mathcal N=2$ supersymmetry tests of appendix~\ref{app:chkCJ3}, which constrain three
linear combinations of the four-loop coefficients, each in its rational and $\zeta_3$ parts.  The $\mathcal N=1$ relation $C_S/C_{S,\rm free}=C_T/C_{T,\rm free}$
(section~\ref{sec:oddsusy}) is a further test, independent of the diagrams, projector and Fourier transform of $\langle TT\rangle$.  At the $S_3$ point it holds
through $\epsilon^3$ without odd loops, and at the other four points with a single odd-loop coefficient; the five points constrain three linear combinations
of the eleven even coefficients of \eqref{eq:CT3}.

\subsection{\texorpdfstring{$C_J$ at $\mathcal O(\epsilon^3)$}{C\_J at O(epsilon\^{}3)}}\label{app:chkCJ3}
At large $N$ we find $C_{J1}=-\frac32\epsilon+\frac12\epsilon^2+\frac{15}{32}\epsilon^3$, the exact Gross--Neveu result \cite{Diab}.  For the $O(n)$ model,
$C_J/C_{J,\rm free}=1-\frac{3(n+2)}{4(n+8)^2}\epsilon^2-\frac{(n+2)(n^2+132n+632)}{8(n+8)^4}\epsilon^3$, identical to the analytic-bootstrap result \cite{DKS,HvL}; its large-$n$ limit agrees with \cite{Petkou}.

The supersymmetry check is more involved.  At the chiral XY point with $N=2$, $C_J^R/C^R_{J,\rm free}=C_T/C_{T,\rm free}$ holds through $\epsilon^3$.
Write $\lambda_{abcd}=\lambda_*(\delta_{ab}\delta_{cd}+\delta_{ac}\delta_{bd}+\delta_{ad}\delta_{bc})$ with $\lambda_*=\frac{\epsilon}{6}+l_2\epsilon^2$, and keep $l_2$ and the
$\mathcal O(\epsilon^3)$ part of $y_*^2$ symbolic.  The latter enters both sides identically, and the $\zeta_3$ terms agree ($\frac{\zeta_3}{36}$ on each side, before the
contribution of $y_*^2$ at $\mathcal O(\epsilon^3)$).  The remaining difference is $\big(\frac49l_2-\frac{2}{81}\big)\epsilon^3$, which vanishes for $l_2=\frac1{18}$,
i.e., for $\lambda_*=y_*^2$ through $\mathcal O(\epsilon^2)$, as supersymmetry requires.  This is the value given by the two-loop $\beta_\lambda$ of \cite{JOS}; our
implementation of it reproduces the GNY and NJLY two-loop $\beta_{g_2}$ of \cite{FGKT}.  With one chiral superfield the check constrains one
linear combination of the four-loop coefficients, in its rational and $\zeta_3$ parts.

We have repeated the test for Wess--Zumino models with several chiral
superfields, all of $R$-charge $\frac23$: $W=XYZ$; $W=\mathrm{Tr}\,\Phi^3$ with $\Phi$ a complex symmetric traceless $3\times3$ matrix; and five random cubic
superpotentials in two and three fields, brought to $\gamma_{ij}\propto\delta_{ij}$ by a $GL(n,\mathbb C)$ field redefinition, which have no symmetry beyond
$U(1)_R$.  We build each theory from its symmetry alone (flavor group, $U(1)_R$ and, where present, complex conjugation), and we solve numerically for
the fixed point with the two-loop $\beta_\lambda$ and the three-loop $\beta_y$, without imposing supersymmetry.  In all eight theories, including $W=Z^3$,
$C_J^R/C^R_{J,\rm free}=C_T/C_{T,\rm free}$ holds through $\epsilon^3$.  Moreover $C_T/C_{T,\rm free}$ takes the same value \eqref{eq:susy} at all
eight fixed points.  This is expected, because in Wess--Zumino models $C_T$ is fixed by the $R$-charges of the chiral superfields, through localization in $d=3$
\cite{ClossetEtAl,NishiokaYonekura} and through the 't~Hooft anomalies in $d=4$ \cite{AFGJ}, and here they are all $\frac23$.  Together these identities
constrain one linear combination of the three-loop coefficients and three linear combinations of the 63 non-zero four-loop coefficients (11 in \eqref{eq:CT3}, 52 in \eqref{eq:CJ3}), each in its rational and $\zeta_3$
parts; the random superpotentials do not increase this number.

The $\mathcal N=1$ flavor-current multiplet gives two further, nearly parallel, constraints at the $O(3)/\mathbb Z_2$ and $SU(3)$ adjoint points
(section~\ref{sec:oddsusy}).

The $\epsilon^{\prime-1}$ terms of the integrals, which determine the $\epsilon^3$ coefficients, are not tested by pole cancellation; they are tested by
the checks above and by the expansion-depth check of section~\ref{sec:renorm}.
Of these, large $N$ probes only the triple-trace structures, the $O(n)$ model only the $\lambda$-only ones, $\mathcal N=2$ supersymmetry three combinations and
the $\mathcal N=1$ relations a few more.  

\section{Details of the loop computation}\label{app:4loop}
\subsection{Four-loop computation}\label{app:4loopsub}
Both correlators are renormalized with the coupling counterterms through two loops,
\[
y_0=\mu^{\epsilon/2}\Big(y+\frac{\beta_1}{\epsilon}+\frac{\beta_2}{2\epsilon}+\frac{D\beta_1[\beta_1]}{2\epsilon^2}\Big),\qquad
\lambda_0=\mu^{\epsilon}\Big(\lambda+\frac{\beta_{\lambda,1}}{\epsilon}\Big).
\]
Here $\beta_L$ and $\beta_{\lambda,L}$ are the $\epsilon$-independent $L$-loop parts of $\beta_y$ and $\beta_\lambda$ in \eqref{eq:rg}, so that
$\beta_y=-\frac\epsilon2y+\beta_1+\beta_2+\dots$ and $\beta_\lambda=-\epsilon\lambda+\beta_{\lambda,1}+\dots$, and $DF[v]=\frac{d}{dt}F(g+tv)\big|_{t=0}$ denotes the
derivative of a function $F$ of the couplings $g=(y,\lambda)$ along $v$; thus $D\beta_1[\beta_1]$ is the derivative of $\beta_1$ along $\beta_1$.  The single
poles are $\beta_L/L$, and the double pole is fixed by them through the consistency of the renormalization group ('t Hooft's pole equations).

At a fixed point $\frac\epsilon2\,y=\beta_1+\beta_2+\dots$ and $\epsilon\lambda=\beta_{\lambda,1}+\dots$.  An invariant $I$ of weight $w(I)$, half its $y$-degree
plus its $\lambda$-degree, satisfies $DI[(\frac12y,\lambda)]=w(I)\,I$, so at a fixed point
\[
\epsilon\,I=\frac{DI[\bar\beta]}{w(I)},\qquad \bar\beta=\big(\beta_1+\beta_2+\dots,\ \beta_{\lambda,1}+\dots\big).
\]
We use this to remove every explicit $\epsilon$ from the coefficients.  Note that $\bar\beta$ is the $\epsilon$-independent part of the $\beta$-functions,
whereas $\beta_y$ in \eqref{eq:rg} is the full $\beta$-function, which vanishes at a fixed point.

The $\langle TT\rangle$ input is the 444 four-loop classes (407 $y^6$, 18 $y^4\lambda$, 16 $y^2\lambda^2$, 3 $\lambda^3$), and the $\langle JJ\rangle$ input the 821 four-loop classes (729 $y^6$, 41 $y^4\lambda$, 41 $y^2\lambda^2$,
10 $\lambda^3$).

We evaluate the flavor factors as in section~\ref{sec:diagrams}, on random theories with an exact $U(1)$ symmetry, and find the basis of
section~\ref{sec:res2}; the repository~\cite{yukawactcj}  lists the candidate invariants and the Ward relations among them.
The four-loop sums $X_{O,k,4}$ for each flavor invariant, in the normalization \eqref{eq:app-forcer-output}, are given in the repository~\cite{yukawactcj} .

Table~\ref{tab:classes} gives the numbers of diagram classes (isomorphism classes of Wick contractions)
at each loop order, with the number of non-vanishing classes in parentheses.  The others vanish identically: 85 four-loop classes
(31 $\langle TT\rangle$, 54 $\langle JJ\rangle$) contain a fermion loop with an odd number of Yukawa vertices, and hence an odd number of $\gamma$ matrices,
which vanishes in our four-dimensional Dirac algebra; in three-dimensional-only theories their contribution is \eqref{eq:oddT} and \eqref{eq:oddJ}
(section~\ref{sec:dirac}), and
22 $\langle JJ\rangle$ classes (2 at three loops, 20 at four) are mapped to themselves by a reflection of the graph that exchanges the two scalar legs of
a $J_\phi$ vertex.  After the transverse projection the $J_\phi$ vertex is odd under this exchange, since the momenta of its legs add up to $Q$, so these
classes vanish.
\begin{table}[h]
\centering\small
\begin{tabular}{lccccc}\toprule
loops & 1 & 2 & 3 & 4 & total\\\midrule
$\langle TT\rangle$ & 2 (2) & 6 (6) & 45 (45) & 444 (413) & 497\\
$\langle JJ\rangle$ & 3 (3) & 11 (11) & 82 (80) & 821 (747) & 917\\\bottomrule
\end{tabular}
\caption{Numbers of diagram classes, with the number of non-vanishing classes in parentheses.}
\label{tab:classes}
\end{table}

\subsection{Fixed points to \texorpdfstring{$\mathcal O(\epsilon^3)$}{O(epsilon\^{}3)}}\label{app:fp3}
At a specific fixed point the $\mathcal O(\epsilon^3)$ terms of $C_T$ and $C_J$ need $y_*^2$ to $\mathcal O(\epsilon^3)$, hence the three-loop $\beta_y$.
We use the general three-loop $\gamma_\phi$, $\gamma_\psi$ and $\tilde\beta_y$ of a scalar--fermion theory as collected in \cite{JOS} (81 coefficients with their tensor structures), in the same flavor-matrix conventions as our calculation.  As in \cite{Zerf}, we do not include the antisymmetric $\upsilon$ terms of \cite{JOS}, which
are relevant only for fixed points of the $B$-function.
With $y_*=\sqrt\epsilon\,(Y_1+\epsilon Y_2+\epsilon^2Y_3)$ and $\lambda_*=\epsilon\,(L_1+\epsilon L_2)$, the corrections $Y_2$, $L_2$ and $Y_3$ follow in turn from linear equations whose matrices are the diagonal blocks of the one-loop stability matrix
(the one-loop $\beta_y$ does not depend on $\lambda$, so the system is triangular);
$L_3$, and therefore the three-loop $\beta_\lambda$, does not enter $C_T$ or $C_J$ at this order.

\section{\texorpdfstring{$C_J$ at $\mathcal O(\epsilon^3)$}{C\_J at O(epsilon\^{}3)}}\label{app:CJ3}
The four-loop terms, which enter at $\mathcal O(\epsilon^3)$ in addition to the higher-order terms of $y_*$ and $\lambda_*$ in \eqref{eq:CJ}, are
\begin{align}\label{eq:CJ3}
\Delta\frac{C_J}{C_{J,\psi}}={}&\big(\tfrac34\zeta_3-\tfrac{9}{32}\big)\Theta \Theta aabbcc+\big(\tfrac92\zeta_3-\tfrac{35}{16}\big)\Theta \Theta aabcbc+\big(\tfrac34\zeta_3-\tfrac58\big)\Theta \Theta aabccb \nonumber\\
&+\big(3\zeta_3-\tfrac52\big)\Theta \Theta abacbc+\big(3\zeta_3-\tfrac{29}{32}\big)\Theta \Theta abaccb-\tfrac1{32}\Theta \Theta abbcac+\tfrac{9}{32}\Theta \Theta abbcca \nonumber\\
&-\big(\tfrac92\zeta_3+\tfrac{11}{2}\big)\Theta \Theta abcabc+\big(\tfrac32\zeta_3-\tfrac{15}{4}\big)\Theta \Theta abcacb+\big(\tfrac32\zeta_3-\tfrac{19}{4}\big)\Theta \Theta abcbac \nonumber\\
&+\tfrac{9}{16}\Theta \Theta abcbca+\big(\tfrac32\zeta_3-\tfrac{27}{16}\big)\Theta \Theta abccab-\tfrac1{32}\Theta \Theta abccba \nonumber\\
&+\tfrac18\Theta a\Theta abbcc+\tfrac{13}{4}\Theta a\Theta abcbc-\tfrac18\Theta a\Theta abccb \nonumber\\
&+\big(\tfrac78-3\zeta_3\big)\Theta a\Theta babcc-2\,\Theta a\Theta bacbc+\big(\tfrac78-3\zeta_3\big)\Theta a\Theta baccb+\tfrac{13}{16}\Theta a\Theta bbacc \nonumber\\
&+\tfrac{13}{2}\Theta a\Theta bcabc+\tfrac{11}{2}\Theta a\Theta bcacb-\tfrac18\Theta aabc\Theta cb-\tfrac{13}{4}\Theta ab\Theta bcac-\tfrac18\Theta ab\Theta bcca \nonumber\\
&-\tfrac52\Theta ab\Theta cabc-\tfrac32\Theta ab\Theta cbac-\tfrac{13}{4}\Theta ab\Theta cbca-\tfrac32\Theta abc\Theta cba \nonumber\\
&+\big(\tfrac34\zeta_3+\tfrac{15}{32}\big)\Theta \Theta aabc\cdot bc+\big(\tfrac34\zeta_3-\tfrac58\big)\Theta \Theta ab\cdot abcc+\big(3\zeta_3-\tfrac{25}{8}\big)\Theta \Theta ab\cdot acbc \nonumber\\
&+\big(\tfrac34\zeta_3-\tfrac58\big)\Theta \Theta ab\cdot accb+\big(\tfrac34\zeta_3+\tfrac{51}{32}\big)\Theta \Theta abac\cdot bc+\tfrac{25}{32}\Theta \Theta abbc\cdot ac \nonumber\\
&-\tfrac{3}{16}\Theta \Theta abca\cdot bc+\big(\tfrac34\zeta_3+\tfrac{51}{32}\big)\Theta \Theta abcb\cdot ac-\tfrac{17}{8}\Theta a\Theta abc\cdot bc \nonumber\\
&-\big(\tfrac32\zeta_3+\tfrac58\big)\Theta a\Theta b\cdot abcc+\big(\tfrac{21}{8}-3\zeta_3\big)\Theta a\Theta b\cdot acbc \nonumber\\
&-\tfrac72\Theta a\Theta bac\cdot bc-\tfrac{13}{8}\Theta ab\cdot \Theta cabc+\tfrac{13}{8}\Theta ab\cdot \Theta cbac \nonumber\\
&+\big(\tfrac34\zeta_3-\tfrac{5}{32}\big)\Theta \Theta ab\cdot ac\cdot bc+\big(\tfrac58-\tfrac34\zeta_3\big)\Theta a\Theta b\cdot ac\cdot bc \nonumber\\
&+9\,\lambda_{abcd}\Theta \Theta abcd-\tfrac{19}{2}\lambda_{abcd}\Theta a\Theta bcd-\tfrac{5}{32}\lambda_{abcd}\lambda_{abce}\Theta \Theta de+\tfrac{19}{48}\lambda_{abcd}\lambda_{abce}\Theta d\Theta e \nonumber\\
&+\tfrac1{16}\theta_{ab}\theta_{ac}\lambda_{bdef}\lambda_{cdeg}\,fg-\tfrac38\theta_{ab}\theta_{ac}\lambda_{bdef}\lambda_{cdgh}\lambda_{efgh} \nonumber\\
&+\tfrac18\theta_{ab}\theta_{cd}\lambda_{acef}\lambda_{bdgh}\lambda_{efgh} \nonumber\\
&-3\,\Theta abc\cdot\Theta abc+(2-6\zeta_3)\,\Theta abc\cdot\Theta bca+\mathcal O(\epsilon^4).
\end{align}
Here a word denotes a trace, $\Theta a\Theta bcabc=\Tr(\Theta y^a\Theta y^by^cy^ay^by^c)$, and a dot separates traces, $\Theta\Theta ab\cdot ac\cdot bc=\Tr(\Theta \Theta y^ay^b)\Tr(y^ay^c)\Tr(y^by^c)$.
All other basis elements have coefficient zero.  The last two terms are \eqref{eq:oddJ}; they vanish for theories obtained from four dimensions.  In \eqref{eq:CJ}, $y_*^2$ is needed to $\mathcal O(\epsilon^3)$ in $\Tr(\Theta \Theta y^ay^a)$, and $\lambda_*$ to $\mathcal O(\epsilon^2)$ in $\theta\,\theta\,\lambda\,\lambda$.

\paragraph{Ward-identity relations.}\label{app:ward}
For any theory obeying \eqref{eq:ward} (verified numerically on $U(1)$ and $SU(2)$ examples):
\begin{align*}
&\Tr(\Theta y^a\Theta y^by^by^a)=\Tr(\Theta y^a\Theta y^ay^by^b),\\ &\Tr(\Theta y^ay^a\Theta y^by^b)=\Tr(\Theta \Theta y^ay^ay^by^b),\\
&\Tr(\Theta y^ay^b\Theta y^ay^b)=2\Tr(\Theta y^a\Theta y^by^ay^b)-\Tr(\Theta \Theta y^ay^by^ay^b),\\
&\Tr(\Theta y^ay^b)\Tr(\Theta y^by^a)=\Tr(\Theta \Theta y^ay^b)\Tr(y^ay^b)-\Tr(\Theta y^a\Theta y^b)\Tr(y^ay^b)\\&\hspace{3cm}+\Tr(\Theta y^ay^b)\Tr(\Theta y^ay^b),\\
&\theta_{ae}\theta_{be}\lambda_{acdf}\lambda_{bcdf}=3\,\theta_{ae}\theta_{bf}\lambda_{abcd}\lambda_{efcd}.
\end{align*}
The first two follow from $[\Theta,y^by^b]=0$.  The last follows from the second line of \eqref{eq:ward}: contracting it with
$\theta_{af}\lambda_{fbcd}$, the first term gives $\theta_{ae}\theta_{be}\lambda_{acdf}\lambda_{bcdf}$ (since $\theta^{\rm T}\theta=\theta\theta^{\rm T}$ for
antisymmetric $\theta$), and each of the other three gives $-\theta_{ae}\theta_{bf}\lambda_{abcd}\lambda_{efcd}$.

\section{Odd fermion loops: the prescription and its alternatives}\label{app:odd}
This appendix gives the details behind section~\ref{sec:dirac}: the continuation of the odd traces, the choices it leaves open, and how the
supersymmetric relations of section~\ref{sec:oddsusy} decide between them.

We use the rules \eqref{eq:tr3} and \eqref{eq:epseps} of section~\ref{sec:dirac}, under which a trace of five $\gamma$'s is \eqref{eq:five} and
depends on where it is opened, \eqref{eq:cyc}.  Multiplying out explicit Pauli matrices gives yet another $D$-dimensional form, identical for
three-dimensional indices and different after integration, so an explicit three-dimensional representation does not decide either.

\paragraph{Diagrams.}  Of the four-loop classes, 31 for $\langle TT\rangle$ and 54 for $\langle JJ\rangle$ contain a fermion loop with an odd
number of Yukawa vertices.  In all but six of each, one odd loop has three Yukawa vertices and no operator.  Its trace depends on two independent
momenta only, so the $\tilde\epsilon$-tensor vanishes after integration.  The remaining six classes have two fermion loops, each with one operator
and three Yukawa vertices, joined by the three scalar lines in the $3!$ possible ways.  For real symmetric $y^a$ all six $\langle TT\rangle$
classes carry $\Tr(y^ay^by^c)\Tr(y^ay^by^c)$.  The $\langle JJ\rangle$ classes carry $\pm I_0$ and $\pm I_1$, with
$I_0=\Tr(\Theta y^ay^by^c)\Tr(\Theta y^ay^by^c)$ and $I_1=\Tr(\Theta y^ay^by^c)\Tr(\Theta y^by^cy^a)$, where the sign changes when one loop
is reversed.  Nothing contributes through three loops, so the results through $\mathcal O(\epsilon^2)$ and the fixed points are unaffected.

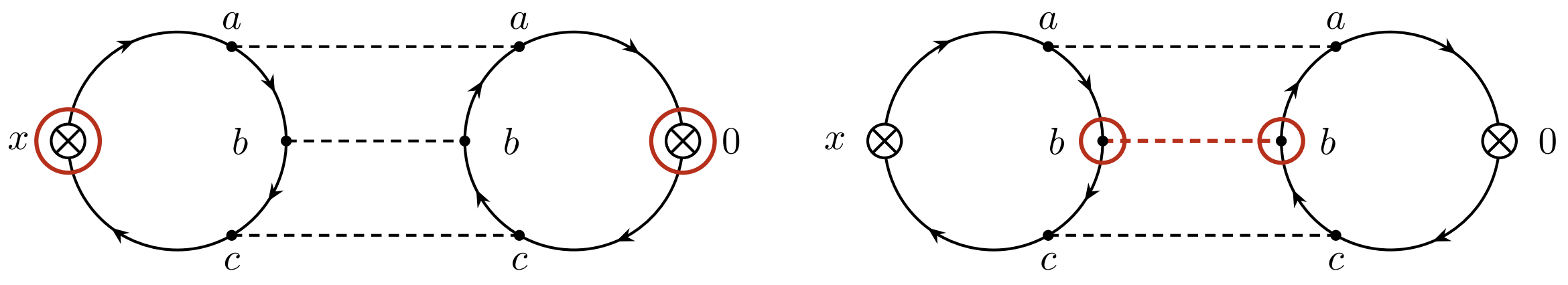
\begin{figure}[htbp]\centering
\newcommand{\oddpair}[1]{%
  \draw[fermion] (-3.1,0) arc (180:60:1.1);
  \draw[fermion] (-1.45,0.953) arc (60:0:1.1);
  \draw[fermion] (-0.9,0) arc (0:-60:1.1);
  \draw[fermion] (-1.45,-0.953) arc (-60:-180:1.1);
  \draw[fermion] (3.1,0) arc (0:-120:1.1);
  \draw[fermion] (1.45,-0.953) arc (240:180:1.1);
  \draw[fermion] (0.9,0) arc (180:120:1.1);
  \draw[fermion] (1.45,0.953) arc (120:0:1.1);
  #1
  \foreach \p in {(-1.45,0.953),(-0.9,0),(-1.45,-0.953),(1.45,0.953),(0.9,0),(1.45,-0.953)} \fill \p circle (0.055);
  \node[font=\small,above=1pt] at (-1.45,0.953) {$a$}; \node[font=\small,left] at (-1.14,0) {$b$}; \node[font=\small,below=1pt] at (-1.45,-0.953) {$c$};
  \node[font=\small,above=1pt] at (1.45,0.953) {$a$}; \node[font=\small,right] at (1.14,0) {$b$}; \node[font=\small,below=1pt] at (1.45,-0.953) {$c$};
  \Tpsi{-3.1}{0} \Tpsi{3.1}{0}
  \node[font=\small,left=7pt] at (-3.1,0) {$x$}; \node[font=\small,right=7pt] at (3.1,0) {$0$};}
\begin{minipage}[b]{0.48\textwidth}\centering\resizebox{\linewidth}{!}{\begin{tikzpicture}
  \useasboundingbox (-3.8,-1.55) rectangle (3.8,1.55);
  \oddpair{\draw[scalar] (-1.45,0.953) to (1.45,0.953); \draw[scalar] (-0.9,0) to (0.9,0); \draw[scalar] (-1.45,-0.953) to (1.45,-0.953);
    \draw[BrickRed,very thick] (-3.1,0) circle (0.32); \draw[BrickRed,very thick] (3.1,0) circle (0.32);}
\end{tikzpicture}}\\[2pt]{\small (a) opened at the operators}\end{minipage}\hfill
\begin{minipage}[b]{0.48\textwidth}\centering\resizebox{\linewidth}{!}{\begin{tikzpicture}
  \useasboundingbox (-3.8,-1.55) rectangle (3.8,1.55);
  \oddpair{\draw[scalar] (-1.45,0.953) to (1.45,0.953); \draw[scalar] (-1.45,-0.953) to (1.45,-0.953);
    \draw[scalar,BrickRed,very thick] (-0.9,0) to (0.9,0);
    \draw[BrickRed,very thick] (-0.9,0) circle (0.22); \draw[BrickRed,very thick] (0.9,0) circle (0.22);}
\end{tikzpicture}}\\[2pt]{\small (b) same-line rule}\end{minipage}
\caption{One of the six four-loop classes with two odd fermion loops; the others differ in how the three scalar lines join the Yukawa vertices
$a,b,c$.  Each loop carries a trace of five $\gamma$-matrices, of the operator and the four fermion propagators, with flavor factor
$\Tr(y^ay^by^c)$ or $\Tr(\Theta y^ay^by^c)$.  Away from $d=3$ this trace is not cyclic, so the result depends on where each trace is opened
(red): at the operator, (a), or at one of the three Yukawa vertices.  Opening at the current violates its Ward identity.  In the same-line rule,
(b), both traces are opened at the two ends of one scalar line; the result is the same for each of the three lines, and it is the
continuation selected by $\mathcal N=1$ supersymmetry (table~\ref{tab:oddcont}).}
\label{fig:odd}
\end{figure}

\paragraph{Where the traces are opened.}  Each loop can be opened at its operator or at one of its three Yukawa vertices
(figure~\ref{fig:odd}).
\begin{itemize}
\item \emph{At the current.}  Inside the current loop the Ward identity reduces $q\!\!\!/$ to a difference of inverse propagators, which needs
cyclicity on one side of the vertex.  Opened at the current, \eqref{eq:five} violates it for $D>3$.  The three-loop subgraph
$\langle J\psi\bar\psi\rangle$ then has a pole, a renormalization of the current by $\tilde\epsilon_{\mu ab}\gamma_a\gamma_b$ that the Ward
identity forbids, and the four-loop $\langle JJ\rangle$ classes have $1/\epsilon'^2$ poles proportional to it.  Opened at a Yukawa vertex, the
identity holds, the subgraph is finite, and every class has a simple pole only.
\item \emph{At the stress tensor.}  The double poles cancel in the sum of the six $\langle TT\rangle$ classes for every choice, but not class by
class.  The simple pole still depends on the choice.
\item \emph{At the same scalar line.}  If both traces are opened at the two ends of one scalar line, each class gives the same result for all three
lines, and the Ward identity of the current holds.  This gives \eqref{eq:oddT} and \eqref{eq:oddJ}.
\end{itemize}
Table~\ref{tab:oddcont} lists the results.  When the same pair of opening points is used in every class, $\kappa_T$ in
$\Delta C_T/C_{T,S}=\kappa_T\Tr(y^ay^by^c)\Tr(y^ay^by^c)$ takes six different values between $-0.89$ and $+1.57$.

\paragraph{Sewing the traces.}  Two further continuations join the two odd traces through one scalar line in ways that are exact in $D=3$.
\begin{itemize}
\item \emph{Fierz.}  For $2\times2$ matrices, $\mathrm{tr}A\,\mathrm{tr}B=\frac12[\mathrm{tr}(AB)+\mathrm{tr}(A\gamma_\mu B\gamma_\mu)]$ merges the
two traces into one even trace, which is unambiguous in $D$ dimensions.  Since $\gamma_\mu\Gamma^{(k)}\gamma_\mu=(-1)^k(D-2k)\Gamma^{(k)}$ for the
antisymmetrized product of $k$ matrices, the grade-three pairing is multiplied by $(7-D)/4$.  Grades one and five give poles in single classes
that cancel between a class and its partner with one loop reversed.
\item \emph{Inversion tensor.}  Pairing the two three-forms through the conformal structure of the scalar line, $x\cdot(2n\wedge(n\lrcorner y)-y)$
with $n$ along the line, equals $x\cdot y$ in $D=3$.  It gives $(4-D)/(D-2)$ times the same-line result, which has no pole, and hence no
$\mathcal O(\epsilon^3)$ term at all.
\end{itemize}
Both reproduce the same-line structure with a factor $f(D)$, $f(3)=1$, and the $\mathcal O(\epsilon^3)$ term scales with $f(4)$.  The opening
points and the normalization $f(4)$ are therefore two independent choices, and all of these continuations agree in $d=3$.

\begin{table}[htbp]
\centering\footnotesize\setlength{\tabcolsep}{4pt}
\begin{tabular}{lccc}\toprule
continuation & $\kappa_T$ & \multicolumn{2}{c}{shift of $\epsilon^3$ in $C_J/C_{J,\rm free}$}\\
 & & $O(3)/\mathbb Z_2$ & $SU(3)$ [susy]\\\midrule
both traces at the operators & $\frac7{24}-\frac{\zeta_3}4=-0.0088$ & \multicolumn{2}{c}{$1/\epsilon'^2$ poles}\\
operator and a Yukawa vertex & $-0.0044$, $-0.1239$ & \multicolumn{2}{c}{$1/\epsilon'^2$ poles}\\
both at Yukawa vertices & $-0.89$, $+0.47$, $+1.57$ & $+0.032$ to $+0.065$ & $+0.036$ to $+0.074$\\
same line, Fierz ($f=\frac{7-D}4$) & $\frac78-\frac34\zeta_3=-0.0265$ & $+0.0897$ & $+0.0961$\\
same line, inversion ($f=\frac{4-D}{D-2}$) & $0$ & $0$ & $0$\\
same line, $\tilde\epsilon\tilde\epsilon=\det[\delta]$ ($f=1$) & $\frac76-\zeta_3=-0.0353902$ & $+0.1196395$ & $+0.1280842$\\\midrule
required by $\mathcal N=1$ supersymmetry & $-0.0353902$ & $+0.1196395$ & $+0.1280842$\\\bottomrule
\end{tabular}
\caption{The odd-loop term for different continuations of the five-$\gamma$ traces.  $\kappa_T$ is the coefficient in
$\Delta C_T/C_{T,S}=\kappa_T\Tr(y^ay^by^c)\Tr(y^ay^by^c)$; for $C_J$ the shift is evaluated at the two $\mathcal N=1$ points of
table~\ref{tab:oddsusy}.  Where the same pair of opening points is used in every diagram, the distinct values are listed.  The last row is
$C_S/C_{S,\rm free}=C_T/C_{T,\rm free}$ at the four $\mathcal N=1$ points with $\Tr(y^ay^by^c)\neq0$ and $C_j/C_{j,\rm free}=C_J/C_{J,\rm free}$
at the two points (table~\ref{tab:oddsusy}).}
\label{tab:oddcont}
\end{table}

\paragraph{The supersymmetric test.}  Since $C_S$ and $C_j$ contain no odd loops at this order (section~\ref{sec:oddsusy}), the relations
$C_S/C_{S,\rm free}=C_T/C_{T,\rm free}$ and $C_j/C_{j,\rm free}=C_J/C_{J,\rm free}$ decide between these continuations without any freedom.  At
each $\mathcal N=1$ point, the difference between $C_S/C_{S,\rm free}$ and $C_T/C_{T,\rm free}$ without odd loops determines $\kappa_T$.  The four
points with $\Tr(y^ay^by^c)\neq0$ give the same value, $\frac76-\zeta_3$, although the ratio of the $\mathcal O(\epsilon)$ coefficient to
$\Tr(y^ay^by^c)\Tr(y^ay^by^c)$ varies from $-24.5$ to $-9.0$ between them.  At the $S_3$ point, where $\Tr(y^ay^by^c)=0$, the two sides agree
exactly; this excludes a relation of the form $C_S/C_{S,\rm free}-1=g(d)\,(C_T/C_{T,\rm free}-1)$ with $g(d)=1+g_2(4-d)^2+\dots$, which would add
$g_2c_1$ at $\mathcal O(\epsilon^3)$ with $c_1=-\frac16$ there.  For $C_J$, the difference at the $O(3)/\mathbb Z_2$ and $SU(3)$ points equals
\eqref{eq:oddJ} with no adjustable parameter; $I_1/I_0$ is $0.90$ at one point and $0.73$ at the other, so both coefficients are tested.  Only the
same-line rule with $f=1$ passes.  It is also the normalization with which the odd-loop terms of the four-loop Yukawa $\beta$-function restore
supersymmetry at the $\mathcal N=1$ point \cite{Zerf}.

\section{The supercurrent and the flavor-current multiplet}\label{app:supercurrent}
For the Wess--Zumino theories of section~\ref{sec:oddsusy}, with $W_a=\partial W/\partial\phi_a=\frac12h_{abc}\phi_b\phi_c$, the supersymmetry
current and its conjugate are, in the conventions of \eqref{eq:action} and appendix~\ref{app:method},
\begin{align}
S_\mu&=(\partial\!\!\!/\phi_a+W_a)\gamma_\mu\psi_a+\xi\,(\gamma_\mu\partial\!\!\!/-\partial_\mu)(\phi_a\psi_a),\nonumber\\
\bar S_\nu&=\bar\psi_a\gamma_\nu(\partial\!\!\!/\phi_a-W_a)+\xi\,\bar\psi_a\phi_a(\overleftarrow{\partial\!\!\!/}\gamma_\nu-\overleftarrow{\partial}_\nu),
\end{align}
with $\xi=(d-2)/(d-1)$.  The opposite signs of the $W$-terms are fixed by the tree-level Ward identity, which fails for the other choice.  On shell,
$\partial^\mu S_\mu=\frac12h_{abc}\,\psi_a(\bar\psi_b\psi_c)$, which vanishes identically for two-component spinors: a cubic in $\psi$ that is totally symmetric in
flavor is totally antisymmetric in three spinor indices that take two values.  The trace is $\gamma^\mu S_\mu=(4-d)\,W_a\psi_a$, the analogue of
$T_\mu{}^\mu\propto(4-d)\lambda\phi^4$.  The lowest component of the flavor-current multiplet is $j=\theta_{ab}\phi_a\psi_b$, with $\bar j=\theta_{ab}\phi_a\bar\psi_b$.

We project $G_{\mu\nu}(Q)=\langle S_\mu\bar S_\nu\rangle$ with $X_S=\mathrm{tr}\big[G_{\mu\nu}\,Q\!\!\!/\,P_{\nu\mu}\big]$,
\begin{equation}
P_{\nu\mu}=\pi_{\nu\mu}-\frac{\gamma^T_\nu\gamma^T_\mu}{d-1},\qquad \pi_{\mu\nu}=\delta_{\mu\nu}-\frac{Q_\mu Q_\nu}{Q^2},\qquad \gamma^T_\mu=\pi_{\mu\rho}\gamma_\rho .
\end{equation}
$P$ is a projector, $P_{\nu\rho}P_{\rho\mu}=P_{\nu\mu}$, with $P_{\mu\mu}=(d-2)\mathbf 1$, and it annihilates $Q_\mu$ and $\gamma_\mu$ from either side.  It therefore
removes the $\xi$-terms and the trace, and also the $W$-terms of $S$ and $\bar S$, which put a bare $\gamma_\mu$ or $\gamma_\nu$ at the ends of the spinor
chain, so that only the $\partial\!\!\!/\phi\,\gamma_\mu\psi$ vertices contribute.  By the same argument, graphs with a cut vertex vanish, as for
$\langle TT\rangle$.  For $j$ we use $X_j=\mathrm{tr}\big[G\,Q\!\!\!/\big]$; $j$ is not conserved, so graphs with a cut vertex are kept.  As for $T$ and $J$, each $L$-loop term is transformed to position space and normalized to the free value.  We read off the
coefficient of $x\!\!\!/\,x_\mu x_\nu$ for $S$, to which only the $Q\!\!\!/\,Q_\mu Q_\nu$ term contributes, so that neither the trace nor longitudinal terms
enter, and the coefficient of $x\!\!\!/$ for $j$.  In the Fourier transform this replaces the factor $s(s+1)(s+2)(s+3)$ of $T$ and $s(s+1)$ of $J$ by
$s(s+1)(s+2)$ and $s$, with $s=d/2+\beta$ for a term $(Q^2)^\beta$.

We generate the diagrams as in section~\ref{sec:method}, with the fermion line open from $\bar S(0)$ to $S(x)$.  At one, two, three and four loops, there are 1, 3, 22
and 222 classes of $\langle S\bar S\rangle$, and 1, 4, 29 and 285 classes of $\langle j\bar j\rangle$.  Those that contain a fermion loop with an
odd number of Yukawa vertices, 2 and 39 at three and four loops for $\langle S\bar S\rangle$ and 2 and 43 for $\langle j\bar j\rangle$, vanish in our Dirac
algebra; by the argument of section~\ref{sec:oddsusy} they would vanish in any continuation.  We checked the Ward identities $Q_\mu G_{\mu\nu}=G_{\mu\nu}Q_\nu=0$, which hold without
contact terms in dimensional regularization, at two and three loops, including the $W$-terms (7 and 54 classes); individual diagrams do not satisfy them.  As for $\langle TT\rangle$, only the couplings are renormalized.  With one superfield, $y=h$ and a general quartic
$\lambda$, the poles of the renormalized $C_S/C_{S,\rm free}$ are
\begin{equation}
-\frac{(3h^2-\lambda)^2}{24\,\epsilon'}\qquad\text{and}\qquad
-\frac{(3h^2-\lambda)^2(11h^2+3\lambda)}{72\,\epsilon^{\prime2}}+\frac{(2h^2+\lambda)(3h^2-\lambda)^2}{48\,\epsilon'}
\end{equation}
at orders $h^4$ and $h^6$.  They vanish on the supersymmetric line $\lambda=3h^2$, and only there.  For several superfields the poles of $C_S$ cancel through
three loops for random $h_{abc}$ with the supersymmetric quartic.  Those of $C_j$ cancel through four loops along the supersymmetric line through the
$O(3)/\mathbb Z_2$ and $SU(3)$ fixed points, so that $j$ needs no renormalization.  At the fixed points we use the couplings of appendix~\ref{app:fp3}.

\bibliographystyle{ssg}
\bibliography{refs}

\end{document}